\documentclass[journal,twocolumn,10pt]{IEEEtran}
\usepackage{textcomp}
\usepackage{stfloats}
\usepackage{url}
\usepackage{graphicx}
\usepackage{verbatim}
\usepackage{upgreek}
\usepackage{amssymb,amsmath}
\usepackage{color}
\usepackage{epstopdf}
\usepackage{bm}
\usepackage{cite}
\usepackage{array,color}
\usepackage{algorithm}
\usepackage{algpseudocode}
\usepackage{amsmath}
\usepackage{amsfonts}
\usepackage{amsthm}
\usepackage{graphics}
\usepackage{epsfig}
\usepackage{soul}
\usepackage{gensymb}
\soulregister\cite7
\soulregister\ref7
\usepackage{subfigure}
\usepackage{setspace}

\newtheorem{rmk}{Remark}

\begin{document}

\title{A Novel Symbol Level Precoding based AFDM Transmission Framework: Offloading Equalization Burden to Transmitter Side}

\author{Shuntian~Tang, Zesong~Fei,~\IEEEmembership{Senior~Member,~IEEE}, Xinyi~Wang,~\IEEEmembership{Member,~IEEE}, Dongkai~Zhou, Zhiqiang~Wei,~\IEEEmembership{Member,~IEEE}, Christos Masouros,~\IEEEmembership{Fellow,~IEEE}

\thanks{A part of this paper was accepted by the 2025 IEEE/CIC International Conference on Communications in China.}

\thanks{Shuntian Tang, Zesong Fei, Xinyi Wang, and Dongkai Zhou are with the School of Information and Electronics, Beijing Institute of Technology, Beijing 100081, China (E-mail:	tst491961069@163.com, feizesong@bit.edu.cn, wangxinyi@bit.edu.cn, bitzdk@163.com).}

\thanks{Zhiqiang Wei is with the School of Mathematics and Statistics, Xi’an Jiaotong University, Xi’an 710049, China, and also with the Peng Cheng Laboratory, Shenzhen, Guangdong 518055, China, and also with the Pazhou Laboratory (Huangpu), Guangzhou, Guangdong 510555, China (email: zhiqiang.wei@xjtu.edu.cn).}

\thanks{Christos Masouros is with the Department of Electronic and Electrical Engineering, University College London, London, UK (e-mail: c.masouros@ucl.ac.uk).}
}



\maketitle

\begin{abstract}
Affine Frequency Division Multiplexing (AFDM) has attracted considerable attention for its robustness to Doppler effects. However, its high receiver-side computational complexity remains a major barrier to practical deployment. To address this, we propose a novel symbol-level precoding (SLP)-based AFDM transmission framework, which shifts the signal processing burden in downlink communications from user side to the base station (BS), enabling direct symbol detection without requiring channel estimation or equalization at the receiver.
Specifically, in the uplink phase, we propose a Sparse Bayesian Learning (SBL) based channel estimation algorithm by exploiting the inherent sparsity of affine frequency (AF) domain channels. In particular, the sparse prior is modeled via a hierarchical Laplace distribution, and parameters are iteratively updated using the Expectation-Maximization (EM) algorithm. We also derive the Bayesian Cramér-Rao Bound (BCRB) to characterize the theoretical performance limit.
In the downlink phase, the BS employs the SLP technology to design the transmitted waveform based on the estimated uplink channel state information (CSI) and channel reciprocity. The resulting optimization problem is formulated as a second-order cone programming (SOCP) problem, and its dual problem is investigated by Lagrangian function and Karush–Kuhn–Tucker conditions. Simulation results demonstrate that the proposed SBL estimator outperforms traditional orthogonal matching pursuit (OMP) in accuracy and robustness to off-grid effects, while the SLP-based waveform design scheme achieves performance comparable to conventional AFDM receivers while significantly reducing the computational complexity at receiver, validating the practicality of our approach.
\end{abstract}

\begin{IEEEkeywords}
Affine frequency division multiplexing (AFDM), Bayesian Cramér-Rao Bound (BCRB), channel estimation, Sparse Bayesian Learning (SBL), symbol-level precoding (SLP).
\end{IEEEkeywords}

\section{Introduction}

\IEEEPARstart{T}{he} sixth generation wireless communication (6G) systems are expected to support a wide range of services and applications, including reliable data transmission at high carrier frequencies under high-mobility conditions~\cite{Saad2020NET}. To meet these demands, novel modulation schemes such as Orthogonal Time Frequency Space (OTFS) and Affine Frequency Division Multiplexing (AFDM) have emerged as promising candidates due to their inherent robustness against Doppler effects. This resilience allows them to outperform conventional Orthogonal Frequency Division Multiplexing (OFDM) in fast time-varying channels~\cite{Rou2024MSP}.

OTFS achieves this robustness by transforming the time-frequency representation of a doubly selective channel into the delay-Doppler (DD) domain, where the channel appears quasi-static, thus simplifying channel estimation and symbol equalization~\cite{Zhou2024ICCC,Zhou2024TVT2,Wei2021MWC}. In contrast, AFDM is a newer modulation scheme inspired by multi-chirp modulation. It employs the inverse discrete affine Fourier transform (IDAFT) along with tunable parameters to achieve effective separation of multipath components in the affine frequency (AF) domain. \textcolor{black}{Moreover, by adjusting these parameters, AFDM offers flexible compatibility with other modulation schemes, such as OFDM, orthogonal chirp division multiplexing (OCDM) and others~\cite{Bemani2023TWC,Cao2024GLOBECOMm,Yin2025arxiv1}}. Both OTFS and AFDM have been shown to offer full diversity gains in doubly selective channels, making them being strong candidates for future high-mobility communication scenarios.

In recent years, growing research interest has been directed toward unlocking the potential of AFDM. In~\cite{Bao2024PIMRC}, the authors investigated how AFDM parameter configurations influence communication performance by affecting diversity gain, thereby providing valuable insights into optimal parameter design. In~\cite{Luo2024TWC}, AFDM was employed to address the challenge of massive connectivity in high-mobility channels, laying a solid foundation for its practical deployment. In~\cite{Rou2025WCL}, the authors explored physical-layer security for AFDM systems, proposing classical and quantum computing-based exhaustive search strategies to effectively prevent eavesdropping. In~\cite{Zhu2024WCL}, the authors proposed a bistatic sensing-assisted channel estimation scheme that leverages AFDM's inherent advantages in both communication and sensing. This work highlights the potential of AFDM in enabling integrated sensing and communication (ISAC) applications. Building on this, the authors in ~\cite{Luo2025IOT} further extended AFDM-ISAC applications to near-field (NF) scenarios, broadening the scope of AFDM’s applicability.

To fully harness the diversity gain of the AFDM technique, accurate channel estimation is essential. In~\cite{Yin2022ICCC}, both single-pilot-aided (SPA) and multiple-pilot-aided (MPA) schemes were investigated, aiming to minimize the mutual interference between pilot and data symbols through careful placement of pilot, guard, and data elements. However, the use of guard symbols significantly reduces spectral efficiency. To address this issue, a guard-free SPA scheme was proposed in~\cite{Xia2025WCL}, where an iterative interference cancellation algorithm was introduced to mitigate pilot-data interference. Furthermore,~\cite{Zheng2025TVT} proposed a channel estimation method based on superimposed pilots to further enhance spectral efficiency, along with an iterative algorithm to eliminate interference between pilot and data symbols. In~\cite{Yin2024TWC}, the authors investigated the widely adopted embedded pilot-aided (EPA) channel estimation scheme in multiple-input multiple-output AFDM (MIMO-AFDM) systems, thereby enhancing AFDM's compatibility with existing MIMO framework. In~\cite{Benzine2024SPAWC}, compressive sensing algorithms were applied to AFDM channel estimation, accompanied by a comprehensive analysis across diverse communication scenarios. Additionally, the authors in ~\cite{Cao2024ICCT} utilized the orthogonal matching pursuit (OMP) algorithm to exploit the inherent sparsity of AFDM channels, thereby enhancing estimation accuracy with reduced complexity.

\textcolor{black}{However, the complexity of the aforementioned channel estimation schemes is typically high, rendering them feasible at the base station (BS) during uplink but impractical at the users with limited computational capabilities during downlink. To address this challenge, a promising approach is to leverage the available channel state information (CSI) at the BS to precode the data symbols prior to transmission. This allows the BS to mitigate the channel effects in advance, thereby significantly reducing the signal processing burden at the user side. The core idea is to suppress inter-symbol interference generated from channel through precoding in advance, akin to how conventional MIMO systems utilize precoding to suppress inter-user interference~\cite{Peel2005TCom,Al-Jarrah2023TWC,Nguyen2019ATC}. Nevertheless, classical block-level precoding (BLP) techniques employed in MIMO precoding typically require the equivalent channel—formed by the product of the precoder and the physical channel—to be equalized at the receiver, which still imposes the burden of CSI acquisition and signal processing at the user side.} 

Compared with conventional BLP schemes that aim to suppress or eliminate interference, symbol level precoding (SLP)  techniques transform interference into a beneficial component and can be applied at the symbol level~\cite{Li2018TWC}. The concept of constructive interference (CI) was first introduced in~\cite{Masouros2007ICC}, where the authors demonstrated that instantaneous interference can be categorized as either constructive or destructive. Building on this idea, the SLP technology was proposed in~\cite{Masouros2015TSP} based on convex optimization. The SLP exploits CSI to pre-compensate transmitted symbols, pushing them away from decision boundaries in the constellation diagram to reduce detection errors. \textcolor{black}{Critically, this implicates that, since the received signal is at the correct constellation region already, thereby eliminating the need for receiver-side CSI acquisition and equalization in constrast to BLP scheme.} Moreover, SLP offers more degrees of freedom (DoFs) in optimization compared to conventional precoding, resulting in improved symbol detection performance~\cite{Alodeh2015TSP}.
Thanks to these advantages, the SLP technique has been extended to various wireless communication scenarios, including millimeter-wave (mmWave) systems~\cite{Liu2019CL}, reconfigurable intelligent surface (RIS)-aided communications~\cite{Liu2022JSTSP}, ISAC~\cite{Liu2018TSP} and multi-user OFDM systems~\cite{Li2025TWC}. Despite its broad applicability, SLP has not yet been explored in AFDM systems, which motivates our work in this paper. 

In this paper, we propose an SLP-based waveform design scheme for AFDM systems to reduce the computational burden on user devices. We first introduce a unified uplink-downlink communication system model and explain how SLP technology can be integrated into AFDM systems. To support waveform design at the BS, we firstly develop a Sparse Bayesian Learning (SBL)-based channel estimation algorithm for the uplink phase, which leverages the inherent sparsity of the AF domain channel. A sparse prior is modeled using a hierarchical structure, and the Expectation-Maximization (EM) algorithm is employed to iteratively update the hyperparameters during the estimation process. We also derive the Bayesian Cramér-Rao lower bound (BCRLB) to theoretically benchmark the performance of the proposed estimator.
With the channel estimate obtained at the BS from the uplink phase, we then propose an SLP-based waveform design scheme for the downlink phase. The waveform design process is first formulated as an optimization problem. To improve efficiency and tractability, we further reformulate the problem using the Lagrangian method, enabling a simplified yet effective solution approach.

The main contributions are summarized as follows:
\begin{itemize}
	\item[$\bullet$] We propose a novel SLP-based user-equalization-free AFDM transmission framework. By leveraging the CSI estimated during the uplink phase, the BS can pre-process the transmitted symbols, allowing user terminals to directly execute symbol detection without the need for additional channel estimation or equalization. This significantly reduces the computational burden at the user side and makes the AFDM system more suitable for resource-constrained user devices.
	\item[$\bullet$] We develop an SBL-based channel estimation scheme tailored for the BS in AFDM systems, which effectively exploits the inherent sparsity of the AF domain channel. The proposed scheme adopts a hierarchical sparse prior and leverages the EM algorithm to iteratively refine the hyperparameters. Furthermore, we derive the BCRLB to serve as a theoretical benchmark for the estimation performance, providing valuable insight into the efficiency of the proposed algorithm.
	\item[$\bullet$] We formulate the downlink SLP-based waveform design process as a constrained optimization problem, which is initially modeled as a second-order cone program (SOCP) problem. To further enhance computational efficiency and enable faster solutions, we reformulate the problem into its Lagrangian dual form and derive a closed-form expression for the precoding symbol vector.
\end{itemize}

\textcolor{black}{The numerical results validate the effectiveness of our proposed user-equalization-free AFDM transmission framework. For the SBL-based channel estimator, we demonstrate its superior robustness to off-grid effects compared to the conventional OMP-based approach. For the SLP-based waveform design scheme, our results show that it achieves performance comparable to the traditional minimum mean square error (MMSE) method under various CSI conditions, confirming its robustness and practical feasibility. Notably, due to its nonlinear nature, the proposed scheme can even outperform the MMSE method in the high SNR regime.}

The remainder of this paper is organized as follows. In section \ref{Preliminaries}, we briefly review AFDM modulation and introduce our AFDM system model. Section \ref{Uplink SBL-Based Channel Estimation} introduces an SBL framework for channel estimation in the uplink phase. In section \ref{Downlink SLP-based Pre-Equalization}, we provide a brief introduction about CI and present the proposed SLP-based waveform design scheme in the downlink phase. Section \ref{Simulation Results} presents simulation results that validate the effectiveness of the proposed transmission scheme. Finally, section \ref{Conclusion} concludes the paper.

\emph{Notations}: Unless otherwise specified, we use a lowercase letter $a$, a boldface lowercase letter $\mathbf{a}$, a boldface capital letter $\mathbf{A}$, and a calligraphy letter $\mathcal{A}$ to denote a scalar, a vector, a matrix, and a set, respectively. $\text{diag}\left(\cdot\right)$ denotes the diagonal operation; $\left(\cdot\right)^*$, $\left(\cdot\right)^T$, and $\left(\cdot\right)^H$ denote the conjugate,  transpose, and Hermitian transpose operations, respectively; $\left(\cdot\right)_N$ denotes the modulo $N$ operation; $\mathbb{R}$ and $\mathbb{C}$ denote the real and complex number fields, respectively; $\text{Re}\{\cdot\}$ and $\text{Im}\{\cdot\}$ denote real and imaginary parts of its argument, respectively; $\mathbf{1}_{N}$ and $\mathbf{I}_{N}$ denote the all-one vector of size $N \times 1$ and identity matrix of size $N \times N$, respectively; \textcolor{black}{$\Gamma^{-1}(\cdot|,a,b)$ and $\Gamma(\cdot|,a,b)$ denote the probability density functions (PDFs) of the inverse Gamma and Gamma distributions with shape parameter $a$ and rate parameter $b$, respectively.
	$\text{Gauss}(\cdot|0, \mathrm{diag}(\bm{\alpha}))$ denotes the PDF of a zero-mean complex Gaussian distribution with covariance matrix $\mathrm{diag}(\bm{\alpha})$}.

\section{Preliminaries}
\label{Preliminaries}
\subsection{Basic Concepts of AFDM}

We first provide an overview of the fundamental principles of AFDM, as introduced in \cite{Bemani2023TWC}. Let $\mathbf{x} = [x_0, x_1, \cdots, x_{N-1}]$ denote an $N \times 1$ complex vector consisting of $Q$-phase shift keying ($Q$-PSK) symbols\footnote{\textcolor{black}{Although this work focuses on PSK modulation, the proposed methods in following sections can be extended to other constellations, such as quadrature amplitude modulation (QAM)~\cite{Li2021TCom}.} \label{s1}}, where $Q$ is the constellation order. By applying the $N$-point IDAFT, $\mathbf{x}$ is transformed from the AF domain to the time domain. The resulting time domain signal is given by
\begin{equation}
	s_n = \frac{1}{\sqrt{N}}\sum_{m=0}^{N-1}{x_m e^{j2\pi\left(c_1n^2+\frac{1}{N}mn+c_2m^2\right)}}, n = 0,1,\cdots,N-1,
\end{equation}
where $c_1$ and $c_2$ are tunable parameters designed to achieve full diversity over doubly selective fading channels.

To preserve signal periodicity in the AF domain, a chirp-periodic prefix (CPP) of length $L_c$ is appended, and is defined as
\begin{equation}
	s_n = s_{N+n}e^{j2\pi c_1\left(N^2+2Nn\right)}, n = -L_c,\cdots,-1,
\end{equation}
where $L_c$ is an integer selected to exceed the maximum delay spread of the wireless channel, measured in samples. Note that a CPP is simply a CP when $2Nc_1$ is an integer value and $N$ is even.

After transmission through a wireless channel with $P$ propagation paths, the received time domain samples $r_n$ can be expressed as
\begin{equation}
	r_n = \sum_{i=1}^{P}{h_i s_{n-l_i}e^{j2\pi f_i n}} + w_n, n = 0,1,\cdots,N-1,
\end{equation}
where $h_i$, $l_i$, and $f_i$ denote the complex channel gain, the normalized time delay (with respect to the sampling period), and the Doppler shift (in digital frequency) of the $i$-th path, respectively. The term $w_n \sim \mathcal{CN}(0, \sigma^2)$ represents additive white Gaussian noise (AWGN) at the receiver.

By applying the $N$-point discrete affine Fourier transform (DAFT), the received signal in the AF domain, denoted by $\mathbf{y} = [y_0, y_1, \cdots, y_{N-1}]$, is obtained as
\begin{align}
	y_m = \frac{1}{\sqrt{N}}&\sum_{n=0}^{N-1}{r_n e^{-j2\pi\left(c_2m^2+\frac{1}{N}mn+c_1n^2\right)}}, \notag \\
	& \qquad\qquad \qquad m = 0,1,\cdots,N-1,
\end{align}

After removing the CPP, the input-output relationship can be expressed in matrix form as
\begin{equation} \label{IO-relation}
	\mathbf{y} =\sum_{i=1}^{P}\mathbf{H}_i\mathbf{x}+\widetilde{\mathbf{w}}= \mathbf{H}_{\text{eff}}\mathbf{x}+\widetilde{\mathbf{w}},
\end{equation}
with 
\begin{equation}
	\mathbf{H}_i=\mathbf{\Lambda}_{c_2}\mathbf{F}\mathbf{\Lambda}_{c_1}\mathbf{\Gamma}_i\mathbf{\Delta}_{f_i}\mathbf{\Pi}^{l_i}\mathbf{\Lambda}_{c_1}^H\mathbf{F}^H\mathbf{\Lambda}_{c_2}^H \in \mathbb{C}^{N \times N},
\end{equation}
where $\widetilde{\mathbf{w}} \sim \mathcal{CN}(0, \sigma^2 \mathbf{I}_N)$ have the same statistics with ${\mathbf{w}}$, $\mathbf{\Lambda}_{c}=\text{diag}\left([1,e^{-j2\pi c},\cdots,e^{-j2\pi c\left(N-1\right)^2}]\right)$, $\mathbf{F}$ is the $N$-point discrete Fourier transform (DFT) matrix, $\mathbf{\Delta}_{f_i} = \text{diag}\left([1,e^{-j2\pi f_i},\cdots,e^{-j2\pi f_i\left(N-1\right)}]\right)$ is the Doppler shift matrix, $\mathbf{\Pi}$ is the permutation (a.k.a.
forward cyclic-shift) matrix, and $\mathbf{\Gamma}_i$ is an $N$-order diagonal matrix given by 
\begin{equation}
	\mathbf{\Gamma}_{i}(i,i)= \begin{cases}e^{-j 2 \pi c_1\left(N^2-2 N\left(l_i-n\right)\right)}, & n<l_i, \\ 1, & n \geq l_i.\end{cases}
\end{equation}
Note that when $2Nc_1$ is an integer value and $N$ is even, $\mathbf{\Gamma}_i=\mathbf{I}_N$.

In (\ref{IO-relation}), for the case of integral normalized Doppler shift, $\mathbf{H}_i$ can be expressed as
\begin{equation} \label{channel_struc}
	\mathbf{H}_i(p, q)= \begin{cases}e^{j \frac{2 \pi}{N}\left(N c_1 l_i^2-q l_i+N c_2\left(q^2-p^2\right)\right)} & q=\left(p+\operatorname{loc}_i\right)_N, \\ 0 & \text { otherwise },\end{cases}
\end{equation}
where $\text{loc}_i = \alpha_i+2Nc_1l_i$ with $\alpha_i=Nf_i \in [-\alpha_{\text{max}}, \alpha_{\text{max}}]$ and $l_i \in [0, l_{\text{max}}]$. The parameters $\alpha_{\text{max}}$ and $l_{\text{max}}$ are defined as the maximum normalized Doppler shift and maximum normalized delay, respectively. As proved in \cite{Bemani2023TWC}, choosing $c_1$ greater than $\frac{2\alpha_{\text{max}}+1}{2N}$ ensures that each row and column of $\mathbf{H}_{\text{eff}}$ contains precisely $P$ non-zero elements.

For the general case that includes fractional normalized Doppler shifts, the channel matrix $\mathbf{H}_i$ can be expressed as
\begin{align}\label{gen_channel_struc}
	\mathbf{H}_i(p, q)&= \frac{1}{N}e^{j\frac{2\pi}{N}\left(Nc_1l_i^2-ql_i+Nc_2\left(q^2-p^2\right)\right)} \notag \\
	&\times \sum_{n=0}^{N-1}e^{-j\frac{2\pi}{N}\left(\left(p-q+v_i+2Nc_1l_i\right)n\right)},
\end{align}
where $v_i = N f_i = \alpha_i + \nu_i$ represents the normalized Doppler shift, with $\alpha_i$ being its integer part and $\nu_i \in [-0.5, 0.5]$ the corresponding fractional part.
\subsection{System Model}

\begin{figure*}[t]
	\centering
	\includegraphics[width=1.05\textwidth]{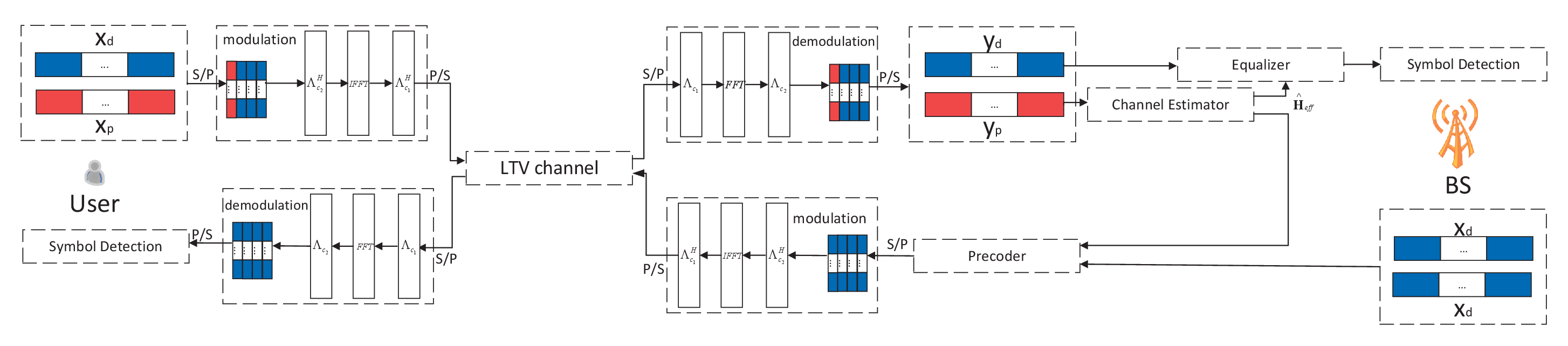}
	\caption{Block diagram of the proposed AFDM transmission framework.}
	\label{fig:sys}
\end{figure*}

In this paper, we propose a BS-exclusive equalization scheme for AFDM systems, as illustrated in Fig. \ref{fig:sys}. The proposed scheme consists of two key phases: \textbf{uplink channel estimation and equalization}, and \textbf{downlink waveform design}.

In the first phase, we assume that each uplink transmission frame remains well within the geometric coherence time, such that the support set, i.e., the indices of the non-zero elements in the $\mathbf{H}_{\text{eff}}$, remains unchanged throughout the frame duration. The transmitted AFDM frame, denoted as $\mathbf{X} \in \mathbb{C}^{N \times L}$, consists of $L$ AFDM symbols per frame, where $LT_s$ is much smaller than the geometric coherence time. Each frame $\mathbf{X}$ includes one pilot symbol $\mathbf{x}_\text{p} \in \mathbb{C}^{N \times 1}$ followed by $L-1$ data symbols $\mathbf{x}_{\text{d},l} \in \mathbb{C}^{N \times 1}$, which can be expressed as
\begin{equation}
	\mathbf{X}= [\mathbf{x}_\text{p}, \mathbf{x}_{\text{d},1},\cdots,\mathbf{x}_{\text{d},L-1}].
\end{equation}
 The first symbol of each frame is reserved for pilot transmission, enabling the BS to estimate the CSI $\tilde{\mathbf{H}}_{\text{eff}}$ for the current transmission. The estimated CSI $\tilde{\mathbf{H}}_{\text{eff}}$ is then used at the BS to equalize the uplink data symbols from the user.

In the second phase, the BS leverages the acquired CSI $\tilde{\mathbf{H}}_{\text{eff}}$ from the first phase to precode the downlink data symbols, exploiting the channel reciprocity between uplink and downlink transmissions\cite{Ran2011ICCTA}. Inspired by \cite{Li2018TWC, Liu2019CL}, the interference and distortions induced by the fast time-varying channel $\mathbf{H}_{\text{eff}}$ are intentionally harnessed as a form of constructive interference. \textcolor{black}{The use of SLP at the BS implicates that interference is constructively exploited such that the received signal is placed at the correct region of the modulation constellation.} As a result, the user can directly detect the data symbols without requiring additional processing, such as channel estimation or equalization. This method is able to significantly reduce the computational burden at the user receiver.

\begin{rmk}
	Due to the high computational complexity of the conventional AFDM receiver and the limited processing capabilities of the user, implementing conventional AFDM scheme in some practical scenarios can be challenging. Our proposed two-phase scheme effectively mitigates this issue by offloading the computational burden to the BS which generally possesses powerful computing ability, making AFDM more feasible for practical deployment.
\end{rmk}

\begin{rmk}
	Moreover, compared with conventional embedded pilot scheme for AFDM\textsuperscript{\cite{Bemani2023TWC}}, our proposed scheme significantly enhances spectral efficiency during the downlink phase. 
\end{rmk}

\section{Uplink SBL-Based Channel Estimator Design}
\label{Uplink SBL-Based Channel Estimation}
In this section, we propose a channel estimation method based on sparse Bayesian learning for the uplink transmission phase. We begin by formulating the sparse channel estimation model based on the input-output relationship in (\ref{IO-relation}). Subsequently, we derive the parameter update equations using Bayesian inference and analyze their computational complexity. Finally, we derive the Bayesian Cramér-Rao Lower Bound (BCRLB) to provide a theoretical performance limit for the proposed SBL-based channel estimation scheme.

\subsection{Sparse Channel Estimation Model for AFDM}
According to (\ref{channel_struc}), the AF domain channel matrix $\mathbf{H}_{\text{eff}}$ exhibits inherent sparsity when the normalized Doppler shifts of all paths are integers. In the case of fractional normalized Doppler shifts, as described in (\ref{gen_channel_struc}), the $\vert \mathbf{H}_i(p, q) \vert$ decreases rapidly as $q$ moves away from $\left(p + \operatorname{loc}_i\right)_N$. Therefore, the channel matrix can still be considered approximately sparse~{\cite{Bemani2023TWC}}.

To leverage this property for channel estimation, we reformulate (\ref{IO-relation}) with respect to the pilot symbol of each frame~{\cite{Bemani2023TWC}}
\begin{equation} \label{liner_IO}
	\mathbf{y}= \sum_{i=1}^{P}\mathbf{H}_i\mathbf{x}_\text{p}+\widetilde{\mathbf{w}} = \mathbf{\Phi}\mathbf{h} + \widetilde{\mathbf{w}},
\end{equation}
\textcolor{black}{where $\mathbf{\Phi}=[\mathbf{H}_1(l_1,v_1)\mathbf{x}_\text{p}|\cdots|\mathbf{H}_M(l_M,v_M)\mathbf{x}_\text{p}] \in \mathbb{C}^{N \times M}$ is the concatenated dictionary matrix  that captures all possible delay-Doppler tap combinations $(l_i,v_i)$, with a total of $M$ atoms, and $\mathbf{h} \in \mathbb{C}^{M \times 1}$ is the corresponding sparse fading vector.}

\textcolor{black}{The dictionary matrix $\mathbf{\Phi}$ defines a subspace of dimension $M$ in which the channel estimation is performed.} The fading vector $\mathbf{h}$ contains the associated coefficients, allowing the channel to be modeled as a linear combination of the atoms in $\mathbf{\Phi}$. Based on (\ref{liner_IO}), the conventional MMSE estimation approach is formulated as
\begin{equation}
	\mathbf{h}_{\text{MMSE}}= \left(\mathbf{\Phi}^H \mathbf{R}_w^{-1} \mathbf{\Phi}+\mathbf{R}_h^{-1}\right)^{-1} \mathbf{\Phi}^H \mathbf{R}_w^{-1} \mathbf{y},
\end{equation}
where $\mathbf{R}_h$ and $\mathbf{R}_w$ denote covariance matrices of $\mathbf{h}$ and $\widetilde{\mathbf{w}}$, respectively. 

Although the conventional MMSE-based channel estimation scheme is widely adopted, {it suffers from high computational complexity, primarily due to multiple matrix inversions. To address this issue, the channel estimation problem can be reformulated as a compressed sensing (CS) task by leveraging the inherent sparsity.} Several classical algorithms, such as OMP\textsuperscript{\cite{Zhang2021ICSPCC}}, can be employed for such sparse signal reconstruction.

However, as a greedy algorithm operating on a discrete grid, conventional OMP is limited to identifying only a few dominant atoms corresponding to on-grid Doppler shifts, resulting in performance degradation under off-grid conditions~{\cite{Wei2022TWC}}. Moreover, under sparsity-uncertain scenarios, OMP may exhibit poor performance or even fail entirely. To overcome these challenges, we adopt the SBL framework for channel estimation, which offers improved accuracy and robustness, particularly in off-grid and sparsity-uncertain environments.

\subsection{SBL-Based Channel Estimation}
In Bayesian modeling, all unknown parameters are treated as random variables with assigned prior probability distributions. In the sparse channel estimation problem formulated in (\ref{liner_IO}), both noise vector $\widetilde{\mathbf{w}}$ and fading vector $\mathbf{h}$ are unknown. Accordingly, we begin by assigning appropriate prior distributions to these variables.

Following the principles of SBL, the noise $\widetilde{\mathbf{w}}$ is modeled as a zero-mean complex Gaussian random vector with unknown variance matrix $\beta \mathbf{I}$, where $\beta$ is an unknown scalar parameter representing the noise precision. To account for the uncertainty in $\beta$, we assign it an inverse Gamma prior distribution.

Specifically, the prior probability density function (PDF) of $\beta$ is denoted by $f\left(\beta;a,b\right)$, and is given by
\begin{equation} \label{prior-distrubution0}
	f\left(\beta;a,b\right)=\Gamma^{-1}(\beta|a,b),
\end{equation}
where $a$ and $b$ are hyperparameters defined by the BS, controlling the shape and rate of the Inv-Gamma distribution, respectively.

Next, we model the fading vector $\mathbf{h}$ using a three-layer hierarchical sparse prior, as proposed in \cite{Babacan2010TIP}, to enable accurate and robust estimation.

In the first layer, $\mathbf{h}$ is assumed to follow a complex Gaussian distribution, with its prior PDF given by
\begin{equation} \label{prior-distrubution1}
	f\left(\mathbf{h};\bm{\alpha}\right)=\text{Gauss}\left(\mathbf{h}|0,\mathrm{diag}\left(\bm{\alpha}\right)\right),
\end{equation}
where $\bm{\alpha} = [\alpha_1, \alpha_2, \dots, \alpha_M]^T$ is a vector of unknown non-negative hyperparameters, where each $\alpha_i$ controls the variance of the corresponding component $h_i$ in $\mathbf{h}$.

In the second layer, to promote sparsity, a Laplace-like prior is imposed on the hyperparameter vector $\bm{\alpha}$ by modeling each element with a Gamma distribution. The corresponding prior PDF is given by
\begin{equation} \label{prior-distrubution2}
	f\left(\bm{\alpha};\lambda\right)=\prod_{i=1}^{M}\Gamma(\alpha_i|1,\frac{\lambda}{2}),
\end{equation}
where $\lambda$ is a unknown hyperparameter that controls the rate of the Gamma distribution.

In the third layer, the hyperparameter $\lambda$ is also treated as a random variable and is assigned a Gamma hyperprior, defined as
\begin{equation} \label{prior-distrubution3}
	f\left(\lambda|\nu\right)=\Gamma(\lambda|\nu/2,\nu/2),
\end{equation}
where $\nu$ is a BS-defined hyperparameter that controls the shape and rate of the Gamma distribution.

To perform Bayesian inference, it is necessary to compute the posterior PDF of the fading vector $\mathbf{h}$.  Specifically, based on the prior distributions given in (\ref{prior-distrubution0})–(\ref{prior-distrubution3}) and by applying Bayes' theorem, the posterior PDF of $\mathbf{h}$ is obtained as~{\cite{Zhao2020CL}}
\begin{equation} 
	f\left(\mathbf{h};\mathbf{y},\bm{\alpha},{\beta},\lambda\right)=\text{Gauss}\left(\mathbf{h}|\bm{\mu},\bm{\Sigma}\right),
\end{equation}
with the posterior mean and covariance matrix given by
\begin{equation} \label{eq:mu_sigma_a}
	\bm{\mu} = \frac{1}{\beta}\bm{\Sigma}\bm{\Phi}^H\mathbf{y},
\end{equation}
and
\begin{equation} \label{eq:mu_sigma_b}
	\bm{\Sigma} = \left(\frac{1}{\beta}\bm{\Phi}^H\bm{\Phi} + \text{diag}(\bm{\alpha})^{-1} \right)^{-1},
\end{equation}
respectively.

We next apply Bayesian inference to optimize the hyperparameters and derive the channel estimate accordingly. Bayesian inference relies on the posterior distribution, which is formulated as
\begin{align}
	p\left(\mathbf{h}|\mathbf{y},\bm{\alpha},\beta,\lambda\right)&=\frac{p\left(\mathbf{y},\mathbf{h},\bm{\alpha},\beta,\lambda\right)}{p\left(\mathbf{y},\bm{\alpha},\beta,\lambda\right)} \notag\\
	&=\frac{p\left(\mathbf{y}|\mathbf{h},\bm{\alpha},\beta,\lambda\right)p\left(\mathbf{h},\bm{\alpha},\beta,\lambda\right)}{p\left(\mathbf{y}|\bm{\alpha},\beta,\lambda\right)p\left(\bm{\alpha},\beta,\lambda\right)} \notag \\
	&=\frac{p\left(\mathbf{y}|\mathbf{h},\bm{\alpha},\beta,\lambda\right)p\left(\mathbf{h}|\bm{\alpha},\beta,\lambda\right)}{p\left(\mathbf{y}|\bm{\alpha},\beta,\lambda\right)},
\end{align}
where $p\left(\mathbf{y}|\bm{\alpha},\beta,\lambda\right)$, referred to as the marginal likelihood or model evidence, involves integration over all latent variables and hyperparameters. It plays a central role in Bayesian inference by quantifying the model’s ability to explain the observed data under the type-II maximum likelihood framework~{\cite{Babacan2010TIP}}. Accordingly, we aim to maximize the corresponding log-marginal likelihood, also known as the log Bayesian evidence, i.e., $\log f(\mathbf{y}; \bm{\alpha}, \beta, \lambda)$, in order to enhance the model's explanatory power for the given observations $\mathbf{y}$. 

However, computing $p\left(\mathbf{y}|\bm{\alpha},\beta,\lambda\right)$ is typically intractable in closed form due to the high-dimensional integration involved, making direct optimization infeasible. To overcome this issue, we employ the EM framework, which enables iterative maximization of the log-evidence. The resulting optimization problem is formulated as follows:
\begin{align}
	\left(\bm{\alpha}^{[p]}, \beta^{[p]}, \lambda^{[p]}\right)&= \notag\\
	&\underset{\bm{\alpha}, \beta, \lambda}{\arg \max } \left\{\log f\left(\mathbf{y}; \bm{\alpha}^{[p-1]},\beta^{[p-1]},\lambda^{[p-1]}\right)\right\},
\end{align}
where $\bm{\alpha}^{[p-1]}$, $\beta^{[p-1]}$ and $\lambda^{[p-1]}$ represent the estimates obtained from the $(p-1)$-th iteration. 

Specifically, in the Expectation-step of the $(p)$-th iteration, we calculate the parameters of posterior distribution based on (\ref{eq:mu_sigma_a}) and (\ref{eq:mu_sigma_b}) with $\bm{\alpha}^{[p-1]}$ and $\beta^{[p-1]}$ to obtain $\bm{\mu}^{[p]}$ and $\bm{\Sigma}^{[p]}$, respectively.

In the subsequent Maximization-step, these posterior statistics $\bm{\mu}^{[p]}$ and $\bm{\Sigma}^{[p]}$ are then used to update the hyperparameters. The update rules for $\bm{\alpha}$, $\beta$ and $\lambda$ are given as follows{\cite{Babacan2010TIP}}:
\begin{equation} \label{update_alpha}
	\alpha_i^{[p]}=\frac{\sqrt{1+4 \lambda^{[p-1]}\left({\Sigma}^{[p]}_{i i}+\vert\mu^{[p]}_i\vert^2\right)}-1}{2 \lambda^{[p-1]}}, \quad i=1, \cdots, M,
\end{equation}
\begin{equation} \label{update_beta}
	\beta^{[p]}=\frac{2 b+\mathbb{E}\left\{\|\mathbf{y}-\mathbf{\Phi} \mathbf{h}\|_2^2\right\}}{2 a-2+N},
\end{equation}
\begin{equation} \label{update_lambda}
	\lambda^{[p]}=\frac{M+\nu/2-1}{\sum_{i=1}^{M}{\alpha_i^{[p]}}/{2}+\nu/2},
\end{equation}
where $\mathbb{E}\left\{\|\mathbf{y}-\mathbf{\Phi} \mathbf{h}\|_2^2\right\}=\beta^{[p-1]}\sum_{i=1}^{M}\left(1-{\Sigma}^{[p]}_{i i}/{\alpha_i^{[p]}}\right)+\|\mathbf{y}-\mathbf{\Phi} \bm{\mu}^{[p]}\|_2^2$ with $\mathbb{E}(\cdot)$ being the expectation function, and $\nu$ is updated by numerically solving the following nonlinear equation
\begin{equation} \label{update_nu}
	\log\frac{\nu}{2}+1-\psi(\nu)+\log \lambda^{[p]} - \lambda^{[p]} =0,
\end{equation}
where $\psi(\cdot)$ being the digamma function. Since this equation does not admit a closed-form solution, a numerical method is employed.

The EM procedure described above is repeated for a maximum of $N_{\text{max}}$ iterations or until the stopping criterion $\Vert \bm{\alpha}^{[p]} - \bm{\alpha}^{[p-1]} \Vert_2^2 \leq \epsilon$ is satisfied. The converged posteriori mean $\bm{\mu}$ is output as the estimation result $\tilde{\mathbf{h}}${\cite{Srivastava2021TVT}}. 

Subsequently, $\tilde{\mathbf{h}}$ is employed to construct the estimated channel matrix $\tilde{\mathbf{H}}_{\text{eff}}$ as follows
\begin{equation} \label{CS_comb}
	\tilde{\mathbf{H}}_{\text{eff}} = \sum_{i=1}^{M}\tilde{{h}}_i\mathbf{H}_i.
\end{equation}

By defining a combination factor $\eta~(\eta > 1)$, an alternative scheme for channel estimation is to construct $\tilde{\mathbf{h}}$ by selecting the first $\eta P$ elements of $\bm{\mu}$ with the largest amplitudes~{\cite{Zhao2020CL}}, where $P < \Delta = \eta P \leq M$. Although each element in the estimated vector $\bm{\mu}$ contains partial channel information, the corresponding SNR varies. By selectively combining only the elements with higher amplitudes, the overall SNR of the reconstructed channel can be enhanced. However, this advantage diminishes at high SNR regimes. In such scenarios, the additional channel information contained in the remaining elements becomes significant. Hence, utilizing all elements in the combination process ultimately leads to superior performance.

In summary, the iterative procedure of the proposed SBL-based channel estimation algorithm is outlined in Algorithm \ref{alg:SBL1}.

\textcolor{black}{In the following, we analyze the computational complexity of the proposed SBL-based approach~\cite{Wei2022TWC}; the convergence performance will be demonstrated through simulation in Section \ref{Simulation Results}.} 

\textcolor{black}{
	In the E-step, the main computational cost lies in computing the posterior covariance matrix $\bm{\Sigma}$ based on~(\ref{eq:mu_sigma_a}) and the posterior mean $\bm{\mu}$ based on~(\ref{eq:mu_sigma_b}). Specifically, the multiplication $\bm{\Phi}^H \bm{\Phi}$ incurs a complexity of $\mathcal{O}(NM^2)$, while inverting an $M \times M$ matrix requires $\mathcal{O}(M^3)$ operations. Calculating $\bm{\mu}$ further involves matrix-vector multiplications, with a total complexity of $\mathcal{O}(NM + M^2)$. Therefore, the overall complexity of the E-step is $\mathcal{O}(NM^2 + M^3)$.
}

\textcolor{black}{
	In the M-step, updating the hyperparameters primarily involves element-wise operations and simple matrix-vector products:
	\begin{itemize}
		\item[$\bullet$] Updating $\bm{\alpha}$ requires $\mathcal{O}(M)$ operations.
		\item[$\bullet$] Updating $\beta$ involves computing the expected reconstruction error, including $\bm{\Phi} \bm{\mu}$ and a weighted trace term, resulting in a complexity of $\mathcal{O}(NM + M)$.
		\item[$\bullet$] Updating $\lambda$ and $\nu$ requires $\mathcal{O}(M)$ operations and solving a scalar nonlinear equation, respectively. The latter contributes negligible cost using standard numerical solvers.
	\end{itemize}
}

\textcolor{black}{
	Hence, the per-iteration complexity of the EM procedure is dominated by the E-step, leading to an overall complexity of
	\begin{equation}
		\mathcal{O}(NM^2 + M^3).
	\end{equation}
}

\textcolor{black}{
	The EM algorithm typically converges in a finite number of iterations $N_{\text{max}}$, yielding a total computational complexity of
	\begin{equation}
		\mathcal{O}(N_{\text{max}}(NM^2 + M^3)).
	\end{equation}
}

\textcolor{black}{
	\begin{rmk}
		Although the per-iteration complexity is relatively high, it remains acceptable for the BS with sufficient computational resources in the uplink pahse. Moreover, as shown in Section \ref{Simulation Results}, the proposed method achieves superior performance, justifying the computational cost.
	\end{rmk}
}

\renewcommand{\algorithmicrequire}{\textbf{Input:}}
\renewcommand{\algorithmicensure}{\textbf{Output:}}
\begin{algorithm}[t]
	\caption{SBL-based AFDM Channel Estimation}
	\label{alg:SBL1}
	\begin{algorithmic}[1]
		\Require Measurement $\mathbf{y}$, dictionary matrix $\bm{\Phi}$, hyperparameters $a$, $b$, and $\nu$, combination length $\Delta$, convergence threshold $\epsilon$, and maximum number of iterations $N_{\text{max}}$.
		\State Initialize ${\bm{\alpha}}^{[0]}$, ${\beta}^{[0]}$, $\nu$ and ${\lambda}^{[0]}$.
		\State Set $\delta$ to a value larger than $\epsilon$.
		\State Set iteration index $n = 1$.
		
		\While {$\delta > \epsilon$ and $n \leq N_{\text{max}}$}
		\State Compute $\bm{\mu}$ and $\bm{\Sigma}$ using ${\bm{\alpha}}^{[p-1]}$ and ${\beta}^{[p-1]}$ based on~(\ref{eq:mu_sigma_a}) and~(\ref{eq:mu_sigma_b}).
		\State Update ${\bm{\alpha}}^{[p]}$ and ${\beta}^{[p]}$ using ${\lambda}^{[p-1]}$ according to~(\ref{update_alpha}) and~(\ref{update_beta}).
		\State Update ${\lambda}^{[p]}$ using ${\nu}$ according to~(\ref{update_lambda}).
		\State Update ${\nu}$ based on the updated ${\lambda}^{[p]}$ according to~(\ref{update_nu}).
		\State Increment $n \leftarrow n + 1$.
		\State Compute $\delta = \Vert \bm{\alpha}^{[p]} - \bm{\alpha}^{[p-1]} \Vert_2^2$.
		\EndWhile
		\State Reconstruct the estimated effective channel matrix $\tilde{\mathbf{H}}_{\text{eff}}$ based on~(\ref{CS_comb}).
		\Ensure Estimated effective channel $\tilde{\mathbf{H}}_{\text{eff}}$.
	\end{algorithmic}
\end{algorithm}
\subsection{BCRB for Proposed Channel Esitimator}
In this subsection, we provide the Bayesian Cramér Rao bound (BCRB) to characterize the estimation efficiency of the proposed channel estimation scheme. 

To this end, the Bayesian Fisher information matrix (BFIM) for the sparse channel vector $\mathbf{h}$ is given as~{\cite{Srivastava2021TVT}}
\begin{align} \label{FIM_h}
	\mathbf{J}_B&= \mathbf{J}_D+\mathbf{J}_P \notag\\
	&=\mathbf{\Phi}^H\mathbf{R}_w^{-1}\mathbf{\Phi}+\mathbf{R}_h^{-1},
\end{align}
where $\mathbf{J}_D$ is the data-based BFIM derived from the likelihood function, and $\mathbf{J}_P$ is the prior information matrix from the prior distribution of $\mathbf{h}$.

Accordingly, the error covariance matrix for the estimated effective channel $\mathbf{H}_{\text{eff}}=\mathbf{\Phi}\mathbf{h}$ is given by
\begin{align} \label{BCRB_H}
	\mathbf{R}_e&= \mathbb{E}\{ \left[\mathbf{\Phi}\left(\tilde{\mathbf{h}} - \mathbf{h}\right)\right]\left[\mathbf{\Phi}\left(\tilde{\mathbf{h}} - \mathbf{h}\right)\right]^H \}\notag\\
	&=\mathbf{\Phi}\mathbb{E}\{ \left(\tilde{\mathbf{h}} - \mathbf{h}\right)\left(\tilde{\mathbf{h}} - \mathbf{h}\right)^H \}\mathbf{\Phi}^H\succeq \mathbf{\Phi}\mathbf{J}_B^{-1}\mathbf{\Phi}^H,
\end{align}
where the inequality follows from the Bayesian Cramér-Rao inequality, indicating that no unbiased Bayesian estimator can achieve a lower error covariance than the inverse of the BFIM. Therefore, the mean squared error (MSE) lower bound BCRLB for the proposed SBL-estimator can be expressed as
\begin{align} \label{BCRLB_H}
    \text{BCRLB} = \text{Tr}\left(\mathbf{R}_e\right)	= \text{Tr}\left(\mathbf{\Phi}\left[\mathbf{\Phi}^H\mathbf{R}_w^{-1}\mathbf{\Phi}+\mathbf{R}_h^{-1}\right]^{-1}\mathbf{\Phi}^H\right).
\end{align}

It is important to note that the BCRLB derived above assumes perfect knowledge of the AF domain channel profile, as it utilizes the true hyperparameter matrix $\mathbf{R}_h$ to establish a lower bound on the MSE.

\section{Downlink SLP-based Waveform Design}
\label{Downlink SLP-based Pre-Equalization}
In this section, we present the SLP-based waveform design scheme applied in the second downlink phase, where the BS performs precoding using the CSI obtained during the first phase, significantly alleviating the computational burden at the user receiver. We begin by introducing the symbol-level precoding model and then adapt it to the AFDM downlink communication model. Subsequently, the formulated problem is transformed into more a feasible form, and the Lagrangian method is introduced and applied{\cite{Li2018TWC,Liu2019CL}}. 

\subsection{Introduction of Constructive Interference}
Due to the effects of Doppler shifts, symbols on different subcarriers may interfere with each other, resulting in inter-carrier interference (ICI), which complicates the accurate detection of the transmitted data. By introducing the SLP technology{\cite{Li2025TWC}}, the ICI can be exploited and transformed into a constructive component, thereby enhancing the transmission reliability.

Specifically, unlike conventional approaches that attempt to eliminate ICI, SLP leverages it to construct a so-called “constructive region”, where interference helps drive the received symbol further from the decision thresholds, thereby enhancing symbol detection accuracy.

In the downlink transmit process, the symbol on the $n$-th subcarrier at the receiver can be expressed as
\begin{equation}
	y_n = \mathbf{h}_n \mathbf{x} + \widetilde{{w}}_n,n=0,\cdots,N-1,
\end{equation}
where $\mathbf{h}_n$ denotes the $n$-th row of channel matrix $\mathbf{H}_{\text{eff}}$, which is obtained during the first phase at the BS. The goal is to correctly detect each received symbol $y_n$ as the intended symbol $s_n$, drawn from an $Q$-PSK constellation set $\mathcal{S}$, i.e.,
\begin{equation}
	s_n \in \mathcal{S} = \{\text{exp}\left(\frac{j\pi\left(2i-1\right)}{Q}\right),i=1,\cdots,Q\}.
\end{equation}

As an illustrative example, consider the target received symbol $s_n=\text{exp}\left(j\pi / 4\right)$. The concept of CI for QPSK constellation is illustrated in Fig. 2 \textcolor{black}{for the constellation quadrant with positive real and imaginary parts, corresponding to the nominal constellation point $s_n=1+j$}. In this case, the noise-free received signal is given by $\mathbf{h}_n\mathbf{x}$, and the decision boundaries for $\text{exp}\left(j\pi / 4\right)$ lie along the positive real and imaginary axes. The constructive interference region associated with $s_n$ is shaded in light blue. As shown, any received signal $y_n$ that lies within the first quadrant will be correctly decoded as $s_n$. To achieve this, the BS should design the proper transmit signal $\mathbf{x}$ such that the noise-free received signal lies within the CI region and is sufficiently far from the non-constructive region. This design approach, aiming to provide robustness against noise and residual interference, is referred to as SLP.

\begin{figure}
	\centering
	\includegraphics[width=0.5\textwidth]{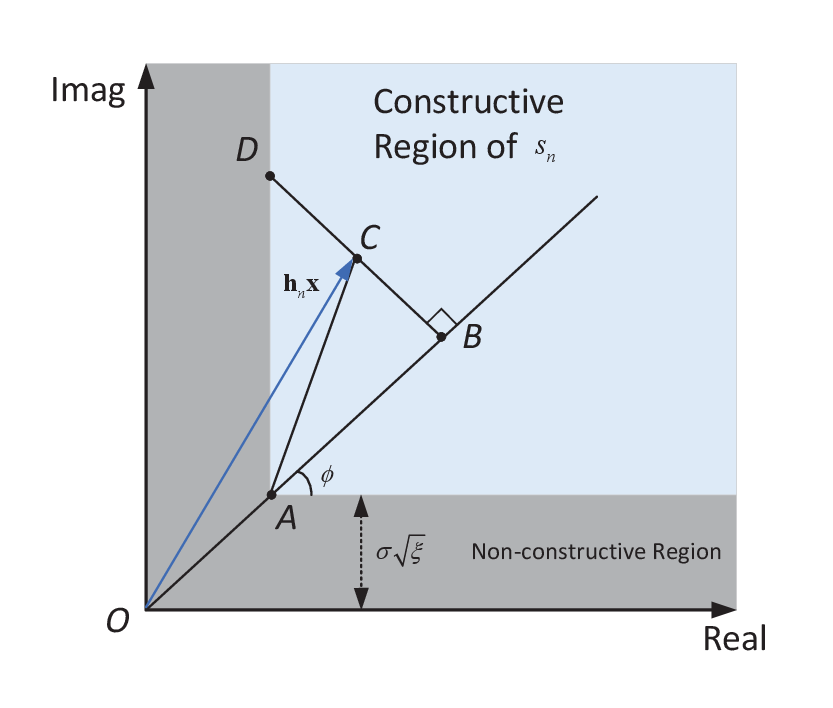}
	\caption{Constructive region schematic diagram for QPSK}
	\label{fig:CI}
\end{figure}

\subsection{SLP-Based Waveform Design for AFDM}
From the mathematical perspective, the construction of CI region is equivalent to ensuring that $|\overrightarrow{B D}|-|\overrightarrow{B C}|\geq0$. To achieve this, we utilize geometric relationships and vector projections. Specifically, we derive the following expression:
\begin{equation}
	|\overrightarrow{B D}|-|\overrightarrow{B C}|=\frac{\text{Re}\{\mathbf{h}_n\mathbf{x}s_n^*-\sigma\sqrt{\xi}\}\tan{\phi}-|\text{Im}\{\mathbf{h}_n\mathbf{x}s_n^*\}|}{\sigma \sqrt{\xi}},
\end{equation}
where $\phi=\frac{2\pi}{Q}$, $\sigma$ denotes the noise power at the receiver, and $\xi$ denotes the signal-to-noise ratio (SNR) of received symbols.

Clearly, a larger value of $|\overrightarrow{B D}|-|\overrightarrow{B C}|$ indicates that the received noise-free signal lies farther away from the non-constructive region, thus improving robustness against additive noise and enhancing the accuracy of symbol detection. Therefore, to achieve optimal symbol detection performance, we aim to maximize the minimum value of $|\overrightarrow{B D}|-|\overrightarrow{B C}|$ across all subcarriers, subject to a total transmit power constraint. The corresponding optimization problem is formulated as follows
\begin{equation} \label{ori_prob}
	\begin{aligned}
		\max_{\mathbf{x}} \quad & \min_n \left\{ \operatorname{Re}\left( \mathbf{h}_n \mathbf{x} s_n^* \right) \sin\phi - \left| \operatorname{Im}\left( \mathbf{h}_n \mathbf{x} s_n^* \right) \right| \cos\phi \right\} \\
		\text{s.t.} \quad & \|\mathbf{x}\|_2^2 \leq P_m,
	\end{aligned}
\end{equation}
where $P_m$ denotes the transmit power budget. To facilitate the solution, we introduce an auxiliary variable $t$, and reformulate the problem (\ref{ori_prob}) as
\begin{equation} \label{prob1}
	\begin{aligned}
		\max_{\mathbf{x},t} \quad & t \\
		\text{s.t.} \quad & \operatorname{Re}\left( \mathbf{h}_n \mathbf{x} s_n^* \right) \sin\phi 
		- \left| \operatorname{Im}\left( \mathbf{h}_n \mathbf{x} s_n^* \right) \right| \cos\phi \geq t, \forall i, \\
		& \|\mathbf{x}\|_2^2 \leq P_m.
	\end{aligned}
\end{equation}

To eliminate the absolute value operator and further simplify the problem (\ref{prob1}), we define the following auxiliary vectors:
\begin{equation} \label{simp1}
	\mathbf{a}_n = \lambda_A\mathbf{h}_ns_n^*,\forall n,
\end{equation}
\begin{equation} \label{simp2}
	\mathbf{b}_n = \lambda_B\mathbf{h}_ns_n^*,\forall n.
\end{equation}
where $\lambda_A=\sin{\phi}-j\cos{\phi}$ and $\lambda_B=\sin{\phi}+j\cos{\phi}$. By using the definitions in (\ref{simp1}) and (\ref{simp2}), the problem (\ref{prob1}) can be reformulated into a more compact and tractable form as follows
\begin{equation} \label{prob2}
	\begin{aligned}
		\max_{\mathbf{x},t} \quad & t \\
		\text{s.t.} \quad & \text{Re}\{\mathbf{A}\mathbf{x}\}\succeq t \cdot \mathbf{1}_N,\\
		& \text{Re}\{\mathbf{B}\mathbf{x}\}\succeq t \cdot \mathbf{1}_N, \\
		& \|\mathbf{x}\|_2^2 \leq P_m,
	\end{aligned}
\end{equation}
where $\mathbf{s}^*=\text{diag}\{s_1^*,\cdots,s_N^*\}$, $\mathbf{A}=\lambda_A \mathbf{s}^* \mathbf{H}_{\text{eff}}$ and $\mathbf{B}=\lambda_B \mathbf{s}^* \mathbf{H}_{\text{eff}}$. 


By defining $\mathbf{w}=[\mathbf{x}_{\text{R}},\mathbf{x}_{\text{I}}]^T \in \mathbb{R}^{2N \times 1}$ where $\mathbf{x}_{\text{R}}$ and $\mathbf{x}_{\text{I}}$ respectively denote the real and imaginary parts of $\mathbf{x}$, and
 $\mathbf{T}=\left[\begin{array}{ll}\mathbf{A}_{\text{R}} & -\mathbf{A}_{\text{I}} \\ \mathbf{B}_{\text{R}} & -\mathbf{B}_\textbf{I}\end{array}\right] \in \mathbb{R}^{2N \times 2N}$ with $\mathbf{A}_{\text{R}}$ and $\mathbf{A}_{\text{I}}$ representing the real and imaginary parts of $\mathbf{A}$, as well as $\mathbf{B}_{\text{R}}$ and $\mathbf{B}_{\text{I}}$ representing the real and imaginary parts of $\mathbf{B}$, the problem (\ref{prob2}) can be further reformulated as follows
\begin{equation} \label{prob3}
	\begin{aligned}
		\max_{\mathbf{w},t} \quad & t \\
		\text{s.t.} \quad & \mathbf{T}\mathbf{w}\succeq t \cdot \mathbf{1}_{2N}, \\
		& \|\mathbf{w}\|_2^2 \leq P_m.
	\end{aligned}
\end{equation}
The problem in (\ref{prob3}) is a SOCP problem, which can be solved using convex optimization tools such as CVX~{\cite{cvx,gb08}}.

To reduce the computational complexity associated with solving the original SOCP problem in (\ref{prob3}), we further analyze it by introducing its corresponding Lagrangian function. This reformulation enables us to derive a more tractable equivalent problem in the dual domain.

Specifically, the Lagrangian function for problem (\ref{prob3}) is expressed as
 \begin{equation} \label{Lagrangian}
	\mathcal{L}\left(t,\mathbf{w},\delta_k\right)=-t+\sum_{k=1}^{2N}\delta_k\left(t-\mathbf{t}_k\mathbf{w}\right)+\delta_0\left(\mathbf{w}^T\mathbf{w}-P_m\right),
 \end{equation} 
where $\mathbf{t}_k$ denotes the $k$-th row of matrix $\mathbf{T}$, and $\delta_k$ $(k=0,1,\cdots,2N)$ are the Lagrange multipliers corresponding to each of the inequality and equality constraints. Based on the Lagrangian function in (\ref{Lagrangian}), the Karush-Kuhn-Tucker (KKT) optimality conditions can be derived as follows
 \begin{equation} \label{KKT1}
	\frac{\partial \mathcal{L}}{\partial t}=-1+\sum_{k=1}^{2N}\delta_k=0,
\end{equation} 
 \begin{equation} \label{KKT2}
	\frac{\partial \mathcal{L}}{\partial \mathbf{w}}=-\sum_{k=1}^{2N}\delta_k\mathbf{t}_k+2\delta_0\mathbf{w}^T=\mathbf{0}_{2N}^T,
\end{equation} 
 \begin{equation} \label{KKT3}
	\delta_k(t-\mathbf{t}_k\mathbf{w})=0,\forall k,
\end{equation} 
 \begin{equation} \label{KKT4}
	\delta_0(\mathbf{w}^T\mathbf{w}-P_m)=0,\forall k,
\end{equation} 

From (\ref{KKT2}), the optimal primal variable $\mathbf{w}$ satisfies
\begin{equation} \label{KKT2_1}
	\mathbf{w}=\frac{\mathbf{T}^T\bm{\delta}}{2\delta_0}.
\end{equation} 

Substituting (\ref{KKT2_1}) into (\ref{KKT4}), the optimal dual variable $\delta_0$ is derived as
\begin{equation} \label{KKT4_1}
	\delta_0=\frac{\Vert\mathbf{T}^T\bm{\delta}\Vert_2}{2\sqrt{P_m}}.
\end{equation} 

By combining (\ref{KKT2_1}) and (\ref{KKT4_1}), the optimal transmit vector $\mathbf{w}$ can be expressed in closed form as
 \begin{equation} \label{wh}
	\mathbf{w}=\sqrt{P_m}\frac{\mathbf{T}^T\bm{\delta}}{\Vert\mathbf{T}^T\bm{\delta}\Vert_2},
\end{equation} 
where $\bm{\delta}=[\delta_1,\cdots,\delta_{2N}]^T \in \mathbb{R}^{2N \times 1}$. 

Moreover, Slater’s condition is satisfied for problem (\ref{prob3}) as shown in \cite{Li2018TWC}, ensuring strong duality. As a result, we can equivalently solve the dual problem to obtain the optimal solution. Based on the stationarity condition in (\ref{KKT1}), we have
\begin{equation} \label{KKT1_1}
	\bm{\delta}^T\mathbf{1}_{2N}=1.
\end{equation} 

Summing all the complementary slackness conditions in (\ref{KKT3}) and applying (\ref{KKT1_1}), we can derive that
\begin{equation} \label{KKT3_1}
	\sum_{k=1}^{2N}\delta_k(t-\mathbf{t}_k\mathbf{w})=0 \Rightarrow t=\bm{\delta}^T\mathbf{T}\mathbf{w}.
\end{equation} 

Substituting the optimal $\mathbf{w}$ in (\ref{wh}) into (\ref{KKT3_1}), we obtain the closed-form optimal $t$ as
\begin{equation} \label{th}
	t = \sqrt{P_m}\Vert\mathbf{T}^T\bm{\delta}\Vert_2.
\end{equation} 

Finally, the optimal dual variable $\bm{\delta}$ can be obtained by solving the following Lagrangian dual problem
\begin{equation} \label{prob4}
	\begin{aligned}
		\min_{\bm{\delta}} \quad & \Vert \mathbf{T}^T\bm{\delta}\Vert_2^2 \\
		\text{s.t.} \quad & \bm{\delta}^T\mathbf{1}_{2N}=1, \\
		& \bm{\delta} \succeq 0.
	\end{aligned}
\end{equation} 

Notably, problem (\ref{prob4}) is a standard quadratic programming (QP) problem over a simplex. It has been demonstrated in \cite{Li2018TWC} that such QP problems can be solved more efficiently than SOCPs of similar scale, using algorithms such as the simplex method or interior-point methods. 

The final transmit symbol vector $\mathbf{x}$ is constructed as
\begin{equation} \label{getx}
	\mathbf{x} = \mathbf{w}[0:N-1] + j\cdot\mathbf{w}[N:2N],
\end{equation}
where the real and imaginary components of $\mathbf{x}$ are formed by splitting $\mathbf{w}$ into its first and second halves, respectively.

In summary, the solution procedure of the SLP-based waveform design algorithm is outlined in Algorithm \ref{alg:SLP}.

\renewcommand{\algorithmicrequire}{\textbf{Input:}}
\renewcommand{\algorithmicensure}{\textbf{Output:}}
\begin{algorithm}[t]
	\caption{SLP-based Waveform Design Algorithm}
	\label{alg:SLP}
	\begin{algorithmic}[1] 
		\Require  downlink symbol vector \( \mathbf{s} \), constellation order $Q$, estimated channel matrix \( \tilde{\mathbf{H}}_{\text{eff}} \), power budget \( P_m \).
		\State Calculate the pre-process channel matrix \( \mathbf{A} \)  and \( \mathbf{B} \) with \( \mathbf{s} \), \( Q \) and $\tilde{\mathbf{H}}_{\text{eff}}$.
		\State Construct  SLP-based channel matrix $\mathbf{T}$ with \( \mathbf{A} \)  and \( \mathbf{B} \).
		\State Solve the problem (\ref{prob4}) with \( \mathbf{T} \), and obtain the dual variable $\bm{\delta}$.
		\State Obtain the optimal transmit vector $\mathbf{w}$ with \( P_m \), $\mathbf{T}$ and $\bm{\delta}$ based on (\ref{wh}).
		\State Construct the precoded symbol vector $\mathbf{x}$ with $\mathbf{w}$ based on (\ref{getx}).
		\Ensure SLP-based precoded symbol vector $\mathbf{x}$.
	\end{algorithmic}
\end{algorithm}

\section{Simulation Results}
\label{Simulation Results}
In this section, we evaluate the effectiveness of the proposed SBL-based channel estimation method and SLP-based waveform design method through Monte Carlo simulations. The simulation parameters are summarized in Table \ref{tab:sim_params}. The Doppler shifts for each path are randomly generated within the range $[-\alpha_{\text{max}}, \alpha_{\text{max}}]$. The Zadoff-Chu (ZC) sequence is employed as the pilot sequence for channel estimation. 
\begin{table}[t]
	\caption{Simulation Parameters}
	\label{tab:sim_params}
	\centering
	\begin{tabular}{|l|l|}
		\hline
		\textbf{Parameter} & \textbf{Value} \\
		\hline
		Subcarrier number & $N = 64$ \\
		Carrier frequency & $f_c = 4~\text{GHz}$ \\
		Subcarrier bandwidth & $\Delta f = 15~\text{kHz}$ \\
		Total number of paths & $P = 3$ \\
		Maximum speed & $v_{\max} = 625~\text{km/h}$ \\
		Maximum delay & $l_{\max} = 2$ \\
		Channel coefficients & $h_i \sim \mathcal{CN}(0, 1/P)$ \\
		\hline
	\end{tabular}
\end{table}

To assess the channel estimation performance, we adopt the normalized mean square error (NMSE), which is defined as
\begin{equation} 
	\text{NMSE}  = 10\log_{10}\mathbb{E}\{\frac{\Vert\tilde{\mathbf{H}}_{\text{es}}-{\mathbf{H}}_{\text{true}}\Vert_F^2}{\Vert{\mathbf{H}}_{\text{true}}\Vert_F^2}\},
\end{equation}
where $\tilde{\mathbf{H}}_{\text{es}}$ denotes the estimated effective channel matrix, and $\mathbf{H}_{\text{true}}$ denotes the real effective channel matrix.

\subsection{SBL-Based Channel Estimation Evaluation}
In this subsection, we present simulation results to evaluate the performance of the proposed SBL-based channel estimation in the uplink phase.

\begin{figure}
	\centering
	\includegraphics[width=0.48\textwidth]{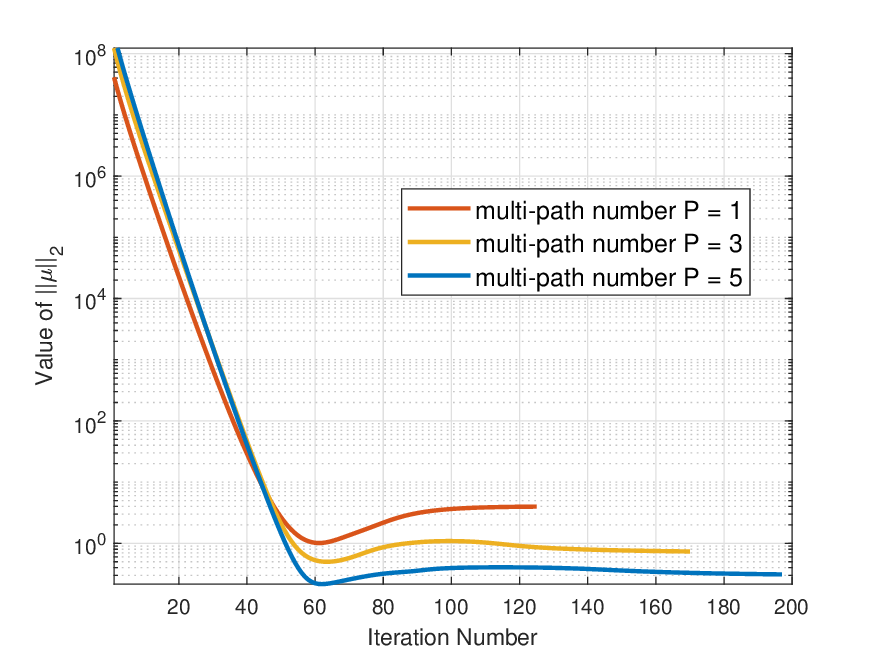}
	\caption{Convergence performance of the SBL-based channel estimation.}\label{fig:1}
\end{figure}

We begin by investigating the convergence behavior of the proposed estimation algorithm. Fig.~\ref{fig:1} depicts the convergence curves of $\Vert\bm{\mu}\Vert_2$ in Algorithm~\ref{alg:SBL1} for different numbers of multi-path $P$. As shown, the algorithm consistently converges within a finite number of iterations across all scenarios, indicating reliable and stable convergence. {It is worth noting that the convergence curves of $\Vert\bm{\mu}\Vert_2$ exhibit a non-monotonic behavior: they initially decline rapidly, followed by a slight increase before reaching a steady state. This phenomenon arises from the iterative activation of relevant components during the hyperparameter update process. In the early iterations, most components are strongly suppressed to enforce channel sparsity, resulting in a sharp drop in $\Vert\bm{\mu}\Vert_2$. As the algorithm progresses, significant channel components begin to emerge and are gradually recovered, leading to a modest rise in the norm, which eventually stabilizes as convergence is achieved.} Moreover, as the number of multi-paths increases, the required number of iterations grows accordingly. This is because a larger $P$ introduces a more intricate sparse structure in the channel, making the recovery of channel coefficients more challenging. As a result, the algorithm needs additional iterations to refine the hyperparameters and meet the convergence criterion.

\begin{figure}
	\centering
	\includegraphics[width=0.48\textwidth]{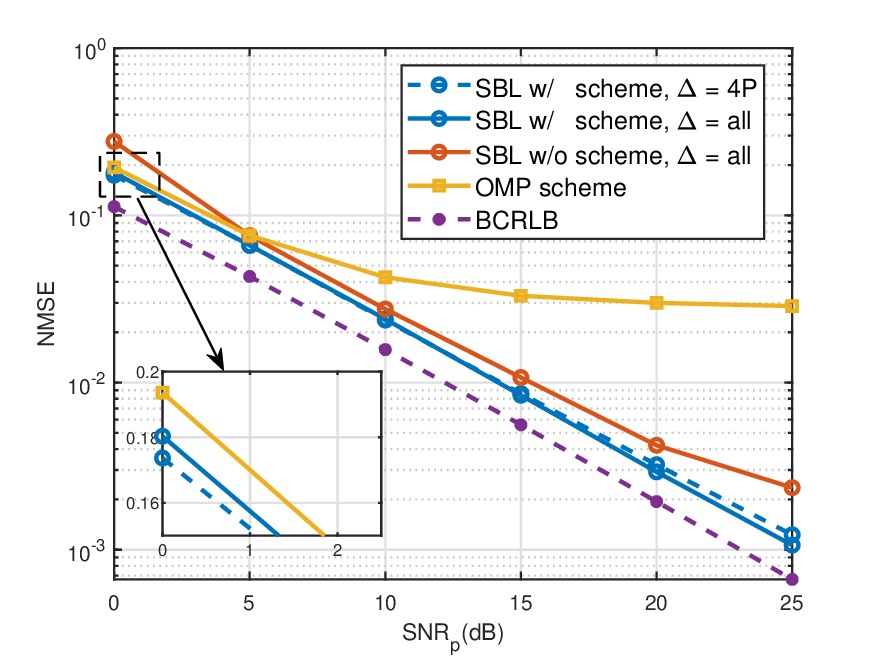}
	\caption{Uplink channel estimation performance under different schemes. ``w/" and ``w/o" denote cases with and without hierarchical Laplace priors, respectively.}\label{fig:2}
 \end{figure}

We then evaluate the NMSE performance of the proposed method in comparison with the OMP scheme, as shown in Fig.~\ref{fig:2}, where ``w/" and ``w/o" refer to SBL-based channel estimation with and without hierarchical Laplace priors, respectively. It is observed that selecting the top $\Delta = 4P$ elements of $\bm{\mu}$ to reconstruct the channel—referred to as partial combination—yields improved NMSE performance in the low-SNR regime (e.g., below 10 dB). This improvement stems from the reduced number of observation paths, which helps suppress noise. However, in the high-SNR regime (e.g., above 10 dB), performance degrades when fewer elements are used, as the residual components contain non-negligible channel information that becomes increasingly important. The gain from this partial combination strategy, as proposed in \cite{Zhao2020CL}, is limited; therefore, it is generally preferable to include all estimated elements when reconstructing the channel matrix.
Moreover, the SBL-based estimation that incorporates hierarchical Laplace priors consistently outperforms its counterpart without such priors. This improvement is attributed to the effective exploitation of the intrinsic sparsity of the AF domain channel, which is better modeled by the Laplace prior. These results highlight the importance of incorporating hierarchical priors in the SBL framework to achieve more accurate channel estimation.
To further benchmark the estimation performance, we also include the BCRLB derived in (\ref{BCRLB_H}) as a theoretical lower bound. The proposed method achieves performance closest to this bound, demonstrating its superior estimation accuracy. 

On the other hand, Fig.~\ref{fig:2} shows that the OMP channel estimation exhibits an error floor. This phenomenon stems from the continuous nature of Doppler shifts. As a greedy algorithm operating on a predefined discrete grid, OMP can only capture a limited number of dominant atoms that align with on-grid Doppler frequencies. When the real Doppler shifts lie off-grid, OMP fails to accurately approximate the true channel parameter, resulting in persistent estimation errors. In contrast, the SBL-based method effectively mitigates off-grid mismatches by leveraging information distributed across multiple components of the posterior mean vector $\bm{\mu}$. Notably, the SBL approach outperforms OMP in the low-SNR regime (e.g., below 5 dB). This performance gain can be attributed to the fact that SBL explicitly models the noise precision and jointly estimates it during the iterative process, thereby exploiting additional noise-related information to enhance channel estimation accuracy.

\begin{figure}
	\centering
	\includegraphics[width=0.48\textwidth]{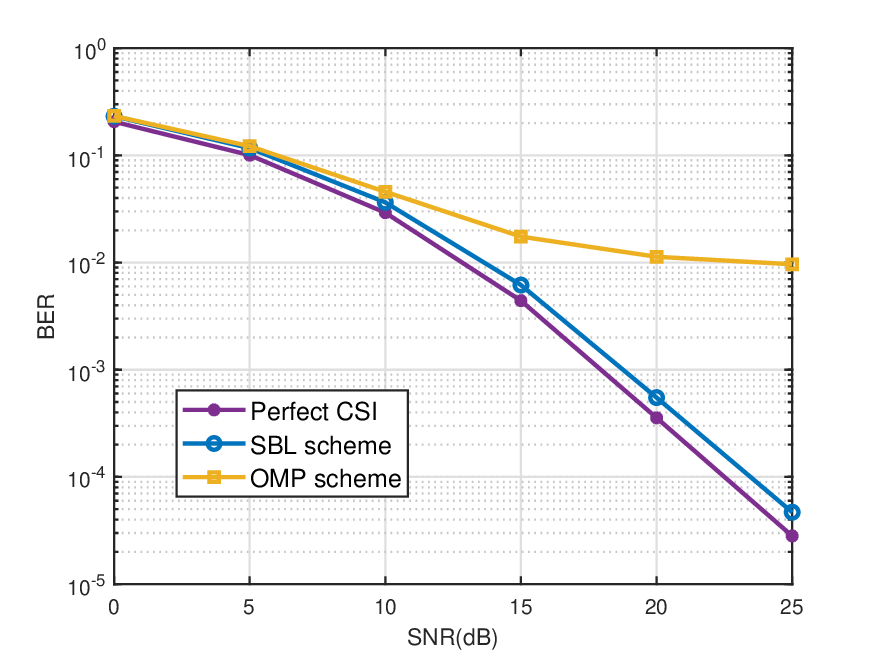}
	\caption{Uplink BER performance under different schemes.}\label{fig:3}
\end{figure}

We further evaluate the impact of the channel estimation results in Fig.~\ref{fig:2} on system performance by conducting channel equalization and analyzing the resulting bit error rate (BER) performance, as depicted in Fig.~\ref{fig:3}. Here, conventional MMSE equalization is applied. It can be observed that the proposed SBL-based channel estimation scheme achieves BER performance that closely approaches the ideal case with perfect CSI. In contrast, the OMP scheme exhibits a significant BER error floor, primarily due to its inability to accurately estimate off-grid channel components, which leads to substantial residual interference during equalization.

\subsection{SLP-Based Waveform Design Evaluation}
In this subsection, we present simulation results to evaluate the effectiveness of the proposed SLP-based waveform design scheme in the downlink phase.

\begin{figure}
	\centering
	\subfigure[]{
		\includegraphics[width=0.225\textwidth]{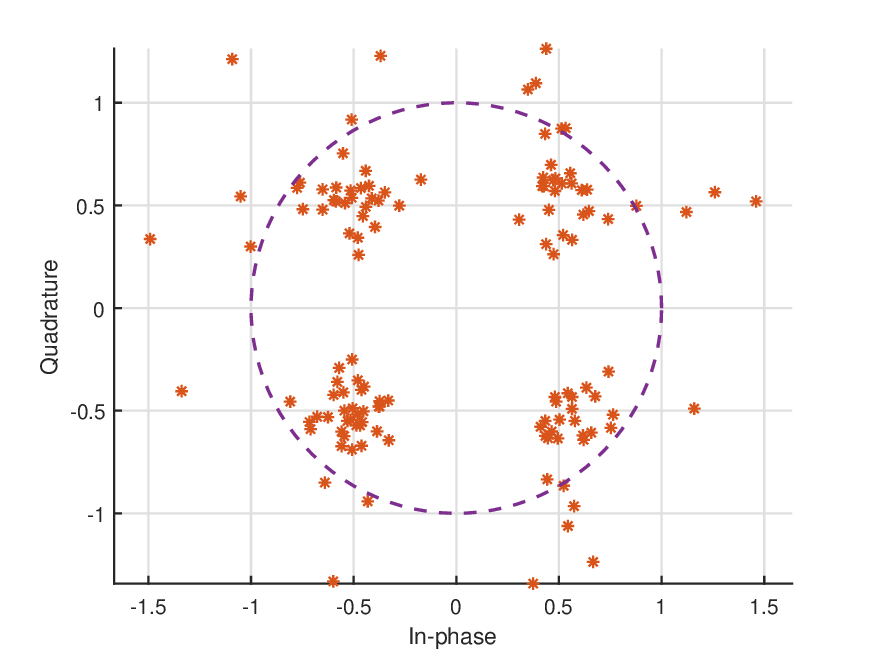}
	}
	\subfigure[]{
		\includegraphics[width=0.225\textwidth]{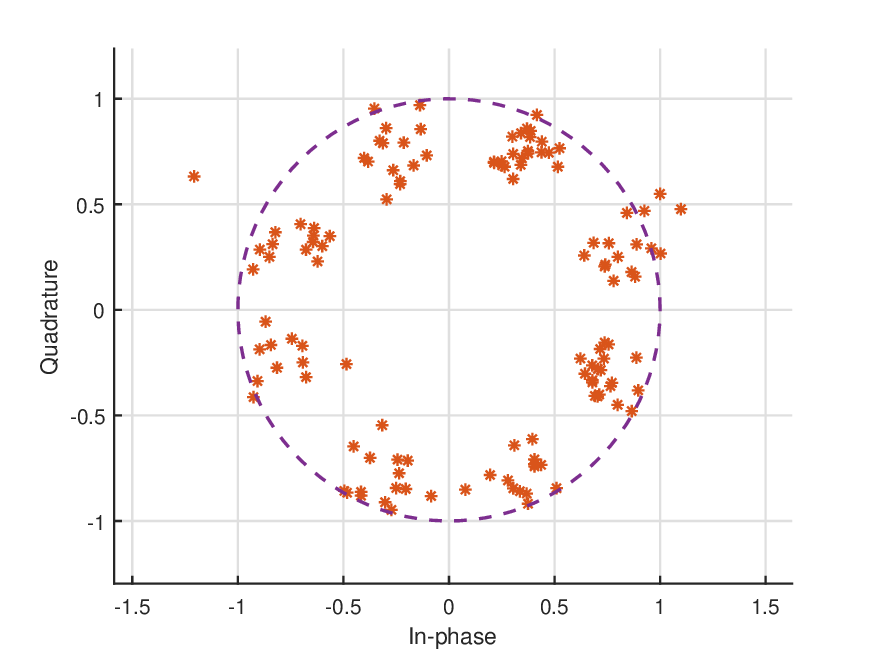}
	}
	\caption{Received symbols constellations after SLP-based precoding: \newline
		\centering (a) QPSK with $Q=4$ (b) 8PSK with $Q=8$}\label{fig:4}

\end{figure}

We first evaluate the performance of the proposed SLP-based waveform design under perfect CSI case. The received symbol constellations at the user side is illustrated in Fig.~\ref{fig:4} to demonstrate the effectiveness of Algorithm~\ref{alg:SLP}. The results are shown for both QPSK and 8PSK modulation with SNR being 20 dB.
The proposed scheme utilizes the CSI to pre-compensate the transmitted symbols such that the received symbols are more likely to fall within their correct decision regions. 
Specifically, in Fig.~\ref{fig:4}(a), for QPSK with $Q = 4$, the received symbols cluster tightly in the first, second, third, and fourth quadrants, indicating that most symbols lie within the correct decision boundaries. Similarly, in Fig.~\ref{fig:4}(b), for 8PSK with $Q = 8$, although the constellation is more densely packed due to increased symbol count, the majority of the received points still reside near the expected 8PSK constellation points along the unit circle. This demonstrates the robustness of the proposed scheme even for higher-order modulations.
These results validate that the SLP-based waveform design effectively compensates for the channel distortion, enabling low-complexity symbol detection at the user side without requiring additional channel equalization.

\begin{figure}
	\centering
	\includegraphics[width=0.48\textwidth]{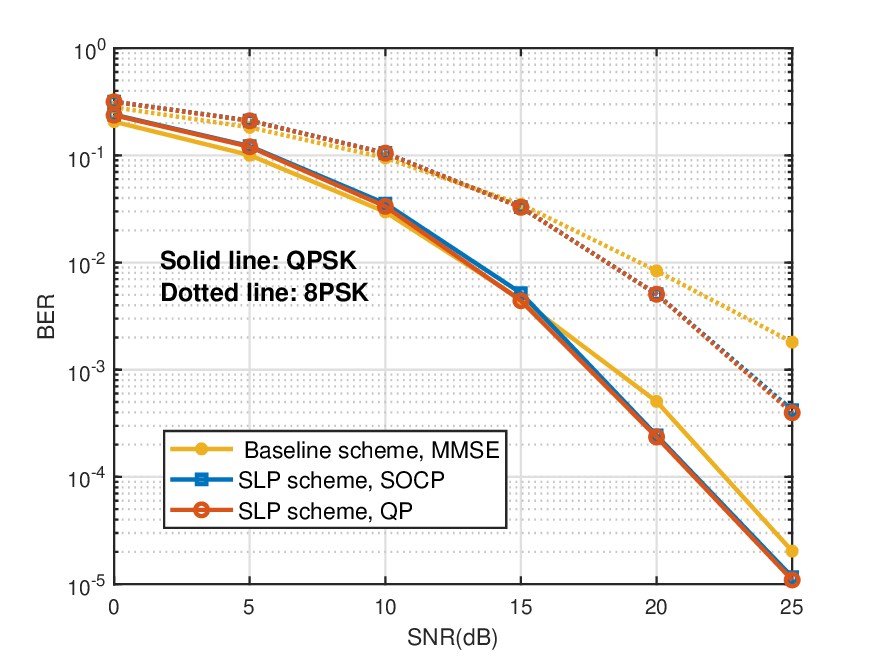}
	\caption{Downlink BER performance under different schemes with perfect CSI.}\label{fig:5}
\end{figure}

As a step further, Fig. \ref{fig:5} illustrates the BER performance of different schemes in the downlink phase for both QPSK and 8PSK modulations.
As observed, the SLP-based schemes (both SOCP and QP formulations) achieve performance comparable to the conventional MMSE scheme and notably surpass it in the high-SNR regime.  This improvement is mainly attributed to the fact that SLP is a nonlinear optimization approach with higher DoFs compared to traditional MMSE equalization. 
Notably, the QP-based solution closely approximates the performance of the SOCP-based one, validating the effectiveness of the proposed Lagrangian-based simplification. According to~\cite{Li2018TWC}, the QP-based method offers improved computational efficiency over SOCP, making it more practical for real-time implementation. 
Moreover, the performance advantage of the SLP scheme becomes more pronounced as the modulation order increases, as seen in the 8PSK results. Although the proposed scheme exhibits slightly inferior performance to MMSE at low SNR, this gap is negligible and does not significantly affect overall system reliability.

\begin{figure}
	\centering
	\includegraphics[width=0.48\textwidth]{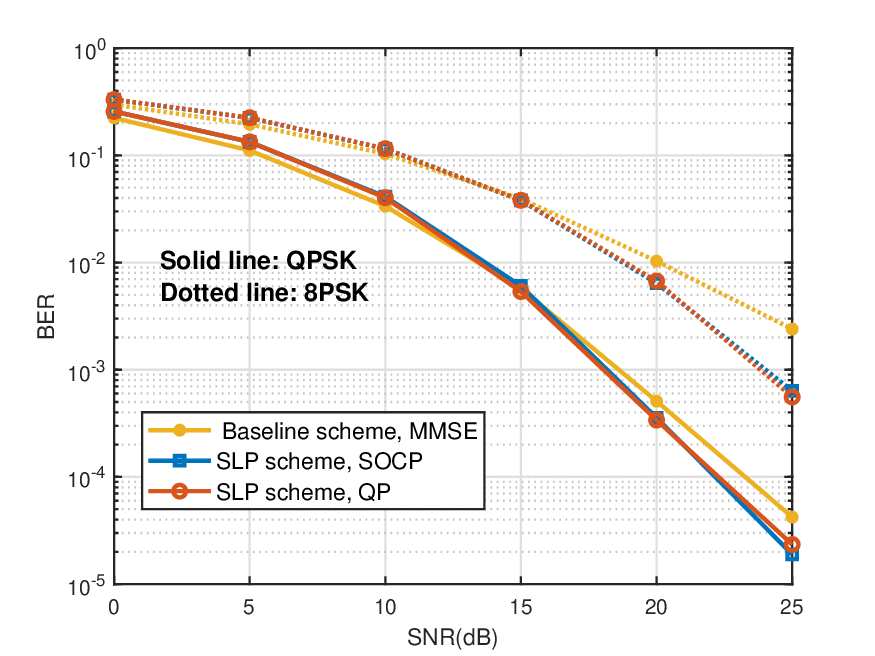}
	\caption{Downlink BER performance under different schemes with estimated CSI.}\label{fig:6}
\end{figure}

To further demonstrate the effectiveness of our proposed scheme, we evaluate its performance using the CSI obtained from the uplink channel estimation phase to precode the downlink transmitted symbols. This setting verifies the practicality of the proposed AFDM system in realistic scenarios. As shown in Fig.~\ref{fig:6}, even when the BS utilizes estimated CSI rather than perfect CSI, the SLP-based schemes still achieve performance comparable to the conventional MMSE scheme. Moreover, the results closely resemble those in Fig.~\ref{fig:5}, where perfect CSI was used, indicating that the proposed SLP-based AFDM transmission framework is both reasonable and feasible.
While a slight performance degradation is observed compared to the perfect CSI case, this loss is expected and remains within an acceptable range, further validating the robustness and practicality of the proposed design in real-world applications.

\begin{figure}
	\centering
	\includegraphics[width=0.48\textwidth]{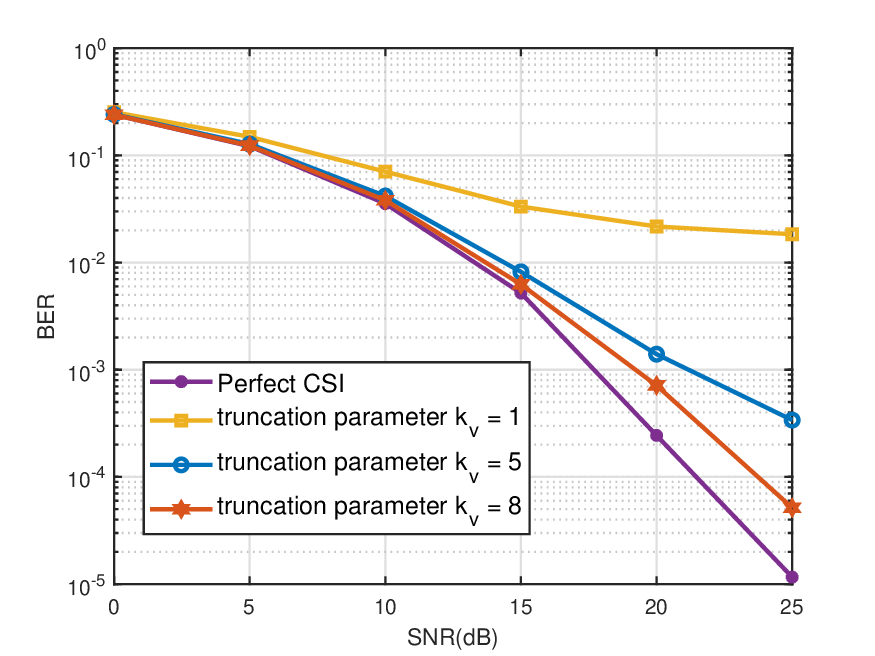}
	\caption{Downlink BER performance under different settings with impaired CSI.}\label{fig:7}
\end{figure}

Finally, we evaluate the impact of impaired channel matrix on the proposed scheme. The generation of the impaired channel follows the approach in \cite{Bemani2023TWC}, where a truncation parameter $k_v$ is introduced to capture the structural sparsity of the channel. The resulting channel matrix is defined as

\begin{equation} \label{channel_struc_error}
	\begin{aligned}
		\mathbf{H}_i(p, q)= & \frac{1}{N} e^{j2 \pi c_1 l_i^2} \sum_{q=\left(p+\operatorname{loc}_i-k_\nu\right)_N}^{\left(p+\operatorname{loc}_i+k_\nu\right)_N} e^{j \frac{2 \pi}{N}\left(-q l_i+N c_2\left(q^2-p^2\right)\right)} \\
		& \times \frac{e^{j2 \pi\left(p-q+a_i+\operatorname{loc}_i\right)}-1}{e^{j \frac{2 \pi}{N}\left(p-q+a_i+\operatorname{loc}_i\right)}-1},
	\end{aligned}
\end{equation}
where all elements of the AF domain channel matrix that lie outside the main peak region are forced to zero, which is determined by the range $k_v$. In other words, the CSI is modeled by truncating the matrix and discarding the insignificant off-peak components. As illustrated in Fig.~\ref{fig:7}, the BER performance of the proposed scheme degrades as off-peak components in the channel matrix are discarded. This highlights the importance of utilizing the complete channel information. It is evident that our scheme relies on the full channel structure to achieve optimal performance, and thus, it is recommended to use the complete channel matrix rather than the sparsely truncated version defined in (\ref{channel_struc_error}).

\section{Conclusion}
\label{Conclusion}
In this paper, we proposed a novel SLP-based AFDM transmission framework, where SLP is incorporated at the BS to perform downlink waveform design, thereby reducing the computational burden on user devices.
For the uplink phase, we introduced an SBL-based channel estimation scheme tailored to AFDM, leveraging the inherent sparsity of AFDM channels. A sparse Bayesian learning framework with hierarchical Laplace priors was adopted, and the EM algorithm was employed to iteratively update the hyperparameters. To benchmark the estimation performance, the BCRLB was derived as a theoretical limit.
In the downlink phase, we proposed an SLP-based waveform design method for AFDM modulation that shifts the receiver processing workload from the user to the BS. The waveform design problem was initially formulated as an SOCP problem and was further reformulated into a Lagrangian dual form to derive a closed-form expression, improving computational efficiency.
Simulation results verified the effectiveness of the proposed framework, demonstrating its potential for practical deployment in AFDM-based high-mobility communication systems.

\bibliographystyle{IEEEtran}%
\bibliography{bib/bibfile}

\begin{thebibliography}{10}
\providecommand{\url}[1]{#1}
\csname url@samestyle\endcsname
\providecommand{\newblock}{\relax}
\providecommand{\bibinfo}[2]{#2}
\providecommand{\BIBentrySTDinterwordspacing}{\spaceskip=0pt\relax}
\providecommand{\BIBentryALTinterwordstretchfactor}{4}
\providecommand{\BIBentryALTinterwordspacing}{\spaceskip=\fontdimen2\font plus
\BIBentryALTinterwordstretchfactor\fontdimen3\font minus
  \fontdimen4\font\relax}
\providecommand{\BIBforeignlanguage}[2]{{%
\expandafter\ifx\csname l@#1\endcsname\relax
\typeout{** WARNING: IEEEtran.bst: No hyphenation pattern has been}%
\typeout{** loaded for the language `#1'. Using the pattern for}%
\typeout{** the default language instead.}%
\else
\language=\csname l@#1\endcsname
\fi
#2}}
\providecommand{\BIBdecl}{\relax}
\BIBdecl

\bibitem{Saad2020NET}
W.~Saad, M.~Bennis, and M.~Chen, ``A vision of {6G} wireless systems:
  Applications, trends, technologies, and open research problems,'' \emph{IEEE
  Net.}, vol.~34, no.~3, pp. 134--142, 2020.

\bibitem{Rou2024MSP}
H.~S. Rou, G.~T.~F. de~Abreu, J.~Choi, D.~González~G., M.~Kountouris, Y.~L.
  Guan, and O.~Gonsa, ``From {Orthogonal Time–Frequency Space to Affine
  Frequency-Division Multiplexing}: A comparative study of next-generation
  waveforms for integrated sensing and communications in doubly dispersive
  channels,'' \emph{IEEE Signal Process. Mag}, vol.~41, no.~5, pp. 71--86,
  2024.

\bibitem{Zhou2024ICCC}
D.~Zhou, S.~Wang, Z.~Zheng, J.~Guo, Z.~Fei, and W.~Yu, ``A cross-domain {PAPR}
  reduction method for {OTFS} modulation,'' in \emph{2024 IEEE/CIC
  International Conference on Communications in China (ICCC)}, 2024, pp.
  2089--2094.

\bibitem{Zhou2024TVT2}
D.~Zhou, J.~Guo, S.~Wang, Z.~Zheng, Z.~Fei, W.~Yuan, and X.~Wang,
  ``{OTFS}-based robust {MMSE} precoding design in over-the-air computation,''
  \emph{IEEE Trans. Veh. Technol.}, vol.~73, no.~9, pp. 13\,932--13\,937, 2024.

\bibitem{Wei2021MWC}
Z.~Wei, W.~Yuan, S.~Li, J.~Yuan, G.~Bharatula, R.~Hadani, and L.~Hanzo,
  ``Orthogonal time-frequency space modulation: A promising next-generation
  waveform,'' \emph{{IEEE} Wireless Commun. Mag.}, vol.~28, no.~4, pp.
  136--144, 2021.

\bibitem{Bemani2023TWC}
A.~Bemani, N.~Ksairi, and M.~Kountouris, ``Affine frequency division
  multiplexing for next generation wireless communications,'' \emph{IEEE Trans.
  Wireless Commun.}, vol.~22, no.~11, pp. 8214--8229, 2023.

\bibitem{Cao2024GLOBECOMm}
R.~Cao, Y.~Zhong, J.~Lyu, D.~Wang, and L.~Fu, ``{AFDM} channel estimation in
  multi-scale multi-lag channels,'' in \emph{GLOBECOM 2024 - 2024 IEEE Global
  Communications Conference}, 2024, pp. 1569--1574.

\bibitem{Yin2025arxiv1}
H.~Yin, Y.~Tang, A.~Bemani, M.~Kountouris, Y.~Zhou, X.~Zhang, Y.~Liu, G.~Chen,
  K.~Yang, F.~Liu, C.~Masouros, S.~Li, G.~Caire, and P.~Xiao, ``Affine
  frequency division multiplexing: Extending {OFDM} for scenario-flexibility
  and resilience,'' 2025, arxiv:2502.04735. [Online]. Available:
  https://arxiv.org/abs/2502.04735.

\bibitem{Bao2024PIMRC}
H.~Bao, H.~Zhuang, Z.~Wang, and G.~Pang, ``Performance trade-off between
  communication and sensing based on {AFDM} parameter adjustment,'' in
  \emph{2024 IEEE 35th International Symposium on Personal, Indoor and Mobile
  Radio Communications (PIMRC)}, 2024, pp. 1--6.

\bibitem{Luo2024TWC}
Q.~Luo, P.~Xiao, Z.~Liu, Z.~Wan, N.~Thomos, Z.~Gao, and Z.~He, ``{AFDM}-{SCMA}:
  A promising waveform for massive connectivity over high mobility channels,''
  \emph{IEEE Trans. Wireless Commun.}, vol.~23, no.~10, pp. 14\,421--14\,436,
  2024.

\bibitem{Rou2025WCL}
H.~S. Rou and G.~T.~F. de~Abreu, ``Chirp-permuted {AFDM} for quantum-resilient
  physical-layer secure communications,'' \emph{IEEE Wireless Commun. Lett.},
  pp. 1--1, 2025.

\bibitem{Zhu2024WCL}
J.~Zhu, Y.~Tang, F.~Liu, X.~Zhang, H.~Yin, and Y.~Zhou, ``{AFDM}-based bistatic
  {Integrated Sensing and Communication} in static scatterer environments,''
  \emph{IEEE Wireless Commun. Lett.}, vol.~13, no.~8, pp. 2245--2249, 2024.

\bibitem{Luo2025IOT}
Y.~Luo, Y.~L. Guan, Y.~Ge, D.~González~G, and C.~Yuen, ``A novel
  angle-delay-doppler estimation scheme for {AFDM-ISAC} system in mixed
  near-field and far-field scenarios,'' \emph{IEEE Internet Things J.},
  vol.~12, no.~13, pp. 22\,669--22\,682, 2025.

\bibitem{Yin2022ICCC}
H.~Yin and Y.~Tang, ``Pilot aided channel estimation for {AFDM} in doubly
  dispersive channels,'' in \emph{2022 IEEE/CIC International Conference on
  Communications in China (ICCC)}, 2022, pp. 308--313.

\bibitem{Xia2025WCL}
H.~Xia, A.~Zhang, D.~Guo, D.~Tian, and S.~Wang, ``A single-pilot-aided channel
  estimation scheme based on interference position indices for {AFDM} in
  delay-doppler channels,'' \emph{IEEE Wireless Commun. Lett.}, pp. 1--1, 2025.

\bibitem{Zheng2025TVT}
K.~Zheng, M.~Wen, T.~Mao, L.~Xiao, and Z.~Wang, ``Channel estimation for {AFDM}
  with superimposed pilots,'' \emph{IEEE Trans. Veh. Technol.}, vol.~74, no.~2,
  pp. 3389--3394, 2025.

\bibitem{Yin2024TWC}
H.~Yin, X.~Wei, Y.~Tang, and K.~Yang, ``Diagonally reconstructed channel
  estimation for {MIMO}-{AFDM} with inter-doppler interference in doubly
  selective channels,'' \emph{IEEE Trans. Wireless Commun.}, vol.~23, no.~10,
  pp. 14\,066--14\,079, 2024.

\bibitem{Benzine2024SPAWC}
W.~Benzine, A.~Bemani, N.~Ksairi, and D.~Slock, ``Affine frequency division
  multiplexing for compressed sensing of time-varying channels,'' in \emph{2024
  IEEE 25th International Workshop on Signal Processing Advances in Wireless
  Communications (SPAWC)}, 2024, pp. 916--920.

\bibitem{Cao2024ICCT}
Z.~Cao, M.~Wen, and Y.~Huang, ``Sparse channel estimation and data detection
  for {AFDM} with superimposed pilot scheme,'' in \emph{2024 IEEE 24th
  International Conference on Communication Technology (ICCT)}, 2024, pp.
  1540--1544.

\bibitem{Peel2005TCom}
C.~Peel, B.~Hochwald, and A.~Swindlehurst, ``A vector-perturbation technique
  for near-capacity multiantenna multiuser communication-part {I}: channel
  inversion and regularization,'' \emph{IEEE Trans. Commun.}, vol.~53, no.~1,
  pp. 195--202, 2005.

\bibitem{Al-Jarrah2023TWC}
M.~Al-Jarrah, E.~Alsusa, and C.~Masouros, ``A unified performance framework for
  integrated sensing-communications based on {KL}20:55 2025/7/2-divergence,''
  \emph{IEEE Trans. Wireless Commun.}, vol.~22, no.~12, pp. 9390--9411, 2023.

\bibitem{Nguyen2019ATC}
H.~T. Nguyen, N.-P. Nguyen, A.~T. Pham, and C.~T. Nguyen, ``Performance
  analysis of ergodic secrecy rates in massive {MIMO-NOMA} networks with
  zero-forcing precoding,'' in \emph{2019 International Conference on Advanced
  Technologies for Communications (ATC)}, 2019, pp. 204--209.

\bibitem{Li2018TWC}
A.~Li and C.~Masouros, ``Interference exploitation precoding made practical:
  Optimal closed-form solutions for psk modulations,'' \emph{IEEE Trans.
  Wireless Commun.}, vol.~17, no.~11, pp. 7661--7676, 2018.

\bibitem{Masouros2007ICC}
C.~Masouros and E.~Alsusa, ``A novel transmitter-based selective-precoding
  technique for {DS/CDMA} systems,'' in \emph{2007 IEEE International
  Conference on Communications}, 2007, pp. 2829--2834.

\bibitem{Masouros2015TSP}
C.~Masouros and G.~Zheng, ``Exploiting known interference as green signal power
  for downlink beamforming optimization,'' \emph{IEEE Trans. Signal Process.},
  vol.~63, no.~14, pp. 3628--3640, 2015.

\bibitem{Alodeh2015TSP}
M.~Alodeh, S.~Chatzinotas, and B.~Ottersten, ``Constructive multiuser
  interference in symbol level precoding for the {MISO} downlink channel,''
  \emph{IEEE Trans. Signal Process.}, vol.~63, no.~9, pp. 2239--2252, 2015.

\bibitem{Liu2019CL}
R.~Liu, H.~Li, and M.~Li, ``Symbol-level hybrid precoding in {mmWave} multiuser
  {MISO} systems,'' \emph{IEEE Commun. Lett.}, vol.~23, no.~9, pp. 1636--1639,
  2019.

\bibitem{Liu2022JSTSP}
R.~Liu, M.~Li, Y.~Liu, Q.~Wu, and Q.~Liu, ``Joint transmit waveform and passive
  beamforming design for {RIS}-aided {DFRC} systems,'' \emph{IEEE J. Special
  Topics Signal Process.}, vol.~16, no.~5, pp. 995--1010, 2022.

\bibitem{Liu2018TSP}
F.~Liu, L.~Zhou, C.~Masouros, A.~Li, W.~Luo, and A.~Petropulu, ``Toward
  dual-functional radar-communication systems: Optimal waveform design,''
  \emph{{IEEE} Trans. Signal Process.}, vol.~66, no.~16, pp. 4264--4279, Aug.
  2018.

\bibitem{Li2025TWC}
P.~Li, M.~Li, R.~Liu, Q.~Liu, and A.~Lee~Swindlehurst, ``{MIMO-OFDM ISAC}
  waveform design for range-doppler sidelobe suppression,'' \emph{IEEE Trans.
  Wireless Commun.}, vol.~24, no.~2, pp. 1001--1015, 2025.

\bibitem{Li2021TCom}
A.~Li, C.~Masouros, B.~Vucetic, Y.~Li, and A.~L. Swindlehurst, ``Interference
  exploitation precoding for multi-level modulations: Closed-form solutions,''
  \emph{IEEE Transactions on Communications}, vol.~69, no.~1, pp. 291--308,
  2021.

\bibitem{Ran2011ICCTA}
J.~Ran and L.~LiICCTA, ``An adaptive method utilizing channel reciprocity in
  {TDD-LTE} system,'' in \emph{IET International Conference on Communication
  Technology and Application (ICCTA 2011)}, 2011, pp. 896--900.

\bibitem{Zhang2021ICSPCC}
S.~Zhang, L.~Xu, and S.~Yan, ``A low complexity {OMP} sparse channel estimation
  algorithm in {OFDM} system,'' in \emph{2021 IEEE International Conference on
  Signal Processing, Communications and Computing (ICSPCC)}, 2021, pp. 1--5.

\bibitem{Wei2022TWC}
Z.~Wei, W.~Yuan, S.~Li, J.~Yuan, and D.~W.~K. Ng, ``Off-grid channel estimation
  with sparse {Bayesian} learning for {OTFS} systems,'' \emph{IEEE Trans.
  Wireless Commun.}, vol.~21, no.~9, pp. 7407--7426, 2022.

\bibitem{Babacan2010TIP}
S.~D. Babacan, R.~Molina, and A.~K. Katsaggelos, ``{Bayesian} compressive
  sensing using laplace priors,'' \emph{IEEE Trans. Image Process.}, vol.~19,
  no.~1, pp. 53--63, 2010.

\bibitem{Zhao2020CL}
L.~Zhao, W.-J. Gao, and W.~Guo, ``Sparse {Bayesian} learning of delay-doppler
  channel for {OTFS} system,'' \emph{IEEE Commun. Lett.}, vol.~24, no.~12, pp.
  2766--2769, 2020.

\bibitem{Srivastava2021TVT}
S.~Srivastava, R.~K. Singh, A.~K. Jagannatham, and L.~Hanzo, ``{Bayesian}
  learning aided sparse channel estimation for orthogonal time frequency space
  modulated systems,'' \emph{IEEE Trans. Veh. Technol.}, vol.~70, no.~8, pp.
  8343--8348, 2021.

\bibitem{cvx}
I.~CVX~Research, ``{CVX}: Matlab software for disciplined convex programming,
  version 2.0,'' \url{https://cvxr.com/cvx}, Aug. 2012.

\bibitem{gb08}
M.~Grant and S.~Boyd, ``Graph implementations for nonsmooth convex programs,''
  in \emph{Recent Advances in Learning and Control}, ser. Lecture Notes in
  Control and Information Sciences, V.~Blondel, S.~Boyd, and H.~Kimura,
  Eds.\hskip 1em plus 0.5em minus 0.4em\relax Springer-Verlag Limited, 2008,
  pp. 95--110, \url{http://stanford.edu/~boyd/graph_dcp.html}.

\end{thebibliography}

\vfill

\end{document}